\documentclass{aa}  
\usepackage{orcidlink}
\usepackage{appendix}
\usepackage{natbib} 
\usepackage{comment}
\usepackage{supertabular}
\usepackage{graphicx}
\usepackage{subcaption}
\usepackage{tablefootnote}
\usepackage{multicol}
\usepackage{pdflscape}
\usepackage{hyperref}
\usepackage{url}
\usepackage{float}
\usepackage{amsmath}
\usepackage{rotating}

\usepackage{txfonts}
\newcommand{\Msun}{M$_{\odot}$}
\newcommand{\Lsun}{L$_{\odot}$}

\newcommand{\Mbd}{M$_{BD}$\ }

\begin{document} 

\title{Complex morphology and kinematics at the heart of the very low luminosity object IRAM 04191+1522
\thanks{This work makes use of archival data obtained with the Spitzer Space Telescope, which was operated by the Jet Propulsion Laboratory, California Institute of Technology under a contract with NASA and the following ALMA data: ADS/JAO.ALMA 2016.1.01284.S and ADS/JAO.ALMA 2019.1.00847.S obtained through the ALMA science archive. ALMA is a partnership of ESO (representing its member states), NSF (USA) and NINS (Japan), together with NRC (Canada), NSTC and ASIAA (Taiwan), and KASI (Republic of Korea), in cooperation with the Republic of Chile. The Joint ALMA Observatory is operated by ESO, AUI/NRAO and NAOJ.}}

\titlerunning{IRAM 04191+1522 Morphology and Kinematics}

\author{N. Otten 
          \inst{1}\thanks{Corresponding author}
          \orcidlink{0000-0002-2530-4137}
          \and
          I. de Gregorio-Monsalvo
          \inst{2}
          \orcidlink{0000-0003-4518-407X}
          \and
          N. Huélamo
          \inst{3}
          \orcidlink{0000-0002-2711-8143}
          \and
          E. Artur de la Villarmois
          \inst{2}
          \orcidlink{0000-0002-8546-9531}
          \and
          G. Blazquez Calero
          \inst{4}
          \orcidlink{0000-0002-7078-2373}
          \and 
          E. Whelan 
          \inst{1}
          \orcidlink{0000-0002-3741-9353}
          }

\institute{Department of Physics, Maynooth University, Maynooth, Co.Kildare, Ireland \\
              \email{noah.otten@mu.ie} 
         \and
             European Southern Observatory, Alonso de Córdova 3107, Vitacura, Casilla 19001, Santiago de Chile, Chile
        \and
            Centro de Astrobiología (CAB), CSIC-INTA, ESAC Campus, Camino bajo del Castillo s/n, E-28692 Villanueva de la Cañada, Madrid, Spain
        \and
            Instituto de Astrofísica de Andalucía Glorieta de la Astronomía s/n, 18008 Granada, Spain }

   \date{Received --; Accepted --}

 
  \abstract
   {The mechanism(s) responsible for the formation of the majority of brown dwarfs (BDs) remains a matter of debate. It is uncertain if the majority of BDs form in molecular cloud cores in a process akin to a scaled-down version of low mass star formation or via disc fragmentation in the circumstellar discs surrounding young stars. To pinpoint these processes, detailed studies of the youngest, most embedded sources must be conducted.}
   {To uncover morphological and kinematic signatures of out-flowing and/or infalling gas structure(s) in the vicinity of the very low luminosity object (VeLLO) IRAM 04191+1522.}
   {Our study utilises archival ALMA sub-mm observations. In particular, we focus this study on the morphological and kinematic analysis of $^{13}$CO~(3-2), C$^{18}$O~(2-1) and SO~6$_{5}$ -- 5$_{4}$ molecular emission lines tracing molecular gas at small spatial scales (few 10's to 100's of au) close to the central source: IRAM 04191+1522. }
   {The red and blueshifted emission traced by $^{13}$CO~(3-2) have very different morphology and kinematics. The blueshifted emission to the north-west of IRAM 04191+1522 may be tracing shocked material in a different direction to the previously reported 0.1 pc ($\sim$ 21000 au) outflow seen in CO. The redshifted emission is mainly seen to the south-east and south-west of IRAM 04191+1522 and could be tracing the base of an outflow cavity. The presence of an outflow cavity in a direction different to any previously reported molecular jet or outflow may suggest the presence of an second outflow in the IRAM 04191+1522 system. The emission traced by C$^{18}$O~(2-1) is highly complex, showing several structures at different spatial scales and separations from the central source. The origin of these components is likely mixed, possibly tracing a combination of emission from a molecular outflow, an outflow cavity and gas in a rotating disc surrounding the central source. The SO~6$_{5}$--5$_{4}$ emission shows evidence of gas in anti-clockwise rotation around the central source. There is also a structure to the north of IRAM 04191+1522 traced in SO~6$_{5}$--5$_{4}$ of undetermined origin.}
   {We detect a new complex set of emission structures traced in $^{13}$CO~(3-2) and C$^{18}$O~(2-1) that indicate the presence of a new outflow cavity, at a different position angle from the previously detected outflow. This new finding reinforces the scenerio that IRAM 04191+1522 is a binary system. The structure to the north of IRAM 04191+1522 detected in SO~6$_{5}$--5$_{4}$ remains of unknown origin and requires further investigation. Follow up observations at high spectral and spatial resolution at intermediate spatial scales are needed to more deeply study these newly detected substructures and their relationship to the previously detected large scale structures. This will allow for a further characterisation of this system and hence a determination of its final fate.}


\maketitle

\section{Introduction}\label{Sec1}
Brown Dwarfs (BDs) are the intermediary mass objects between low mass stars and giant planets. They are principally different objects compared to stars due to their low mass (\Mbd $<$ 0.075~\Msun) and their inability to maintain the consistent fusion of hydrogen \citep{Chabrier2023}. The dominant formation mechanism of these objects remains uncertain. The current literature divides the possible formation processes of BDs into two main branches: star-like or planet-like \citep[see][and references therein]{Palau2024}. In the planet-like scenario, the brown dwarf (BD) forms via turbulent fragmentation of material within a circumstellar disc surrounding a young protostar. This is followed by the subsequent ejection of the BD from the circumstellar disc via multiple body-body interactions \citep{Reipurth2015}. In the star-like scenario, BDs form via the gravitational collapse of fragmented  dense cores of gas and dust within molecular clouds. Turbulence in these cores allows for the production of extreme density fluctuations within the core, decreasing the typical Jeans mass needed for the production of BDs within molecular clouds \citep{PadoanANDNordlund2002,PadoanANDNordlund2004}. Therefore, in order to determine the main formation mechanism of BDs, dedicated studies of the youngest and most embedded objects have to be conducted. In this article we focus on one such young and heavily embedded object, IRAM 04191+1522. 
\newline\par\noindent
Our target IRAM 04191+1522 (hereafter referred to as IRAM04191) is a young ($\approx \times$ 10$^{4}$ years), highly embedded Class 0 system located in Taurus (distance $\sim$ 140 pc) with an estimated protostellar mass of 0.05~\Msun~surrounded by an envelope of mass 0.5~\Msun~(see Table \ref{properties table} and references therein). It was  discovered in 1999 via the detection of a large scale CO (2--1) outflow (0.1 pc) in the southern part of the Taurus star forming region using the IRAM 30m telescope during a survey searching for embedded young stellar objects (YSOs) \citep{Andre1999}. This is particularly interesting as the length scale of this molecular outflow is somewhat comparable to the parsec scale molecular outflows typically driven by Class 0 protostars \citep[e.g.][]{Eisloffel2000}. Since its discovery, IRAM04191 has become a well studied object over the last two decades. Different works have further characterized IRAM04191, and we summarize their main findings below. Following the discovery of IRAM04191, a study of the structure and kinematics of the envelope surrounding this source was conducted by \cite{Belloche2002}. They find evidence of simultaneous infall, outflow and a differential rotation signature between the inner and outer envelope in several molecular species. 
\newline\par\noindent
The first dedicated molecular line study concerning IRAM04191 was conducted by \cite{CFLee2002,CFLee20052005} and mapped the region surrounding IRAM04191 in the CO (1-0), HCO$^{+}$ (1-0) and N$_{2}$H$^{+}$ emission using the Berkeley-Illinois-Maryland Association Array (BIMA) at Hat Creek California. \cite{CFLee20052005} found that the CO (1-0) emission traces two structures along the outflow axis (axis A1 in their Fig. 1, position angle (PA) = 30$^{\circ}$)\footnote{Measured east of north.}. \begin{enumerate}
    \item A bipolar outflow in CO (1--0) along the north-east (red lobe) south-west (blue lobe) direction.
    \item Strong emission $\sim$ 20$\arcsec$ (2800 au) to the north-east of the IRAM04191 source is detected, which they suggest is associated to an internal bow shock driven by an episodic collimated jet.
\end{enumerate} 
\cite{CFLee20052005} also draw a second axis (axis A2 in their Fig. 1, PA = 18$^{\circ}$) from the cavity tip to the centre of the molecular outflow. The authors suggest that this morphology may hint toward a binary nature for IRAM04191. A dense condensation of material in HCO$^{+}$ was also detected (labelled B2 in their Fig. 1) at a distance 40$\arcsec$ (5600 au) to the south-west of IRAM04191. The authors state that this is likely linked to an unresolved expanding shell of gas interacting with the CO outflow. This condensation has also been detected in CS and in CH$_{3}$OH \citep{Takakuwa_2003}. The HCO$^{+}$ emission was also found to trace a structure of $\sim$ 50$\arcsec$ (7000 au) in length at a low velocity of -1 km.s$^{-1}$ to the south-west of IRAM04191 which may be produced via an interaction with the molecular jet. The detected N$_{2}$H$^{+}$ emission was attributed to clumpy dense core of material with a  mean radius of 10$\arcsec$ (1400 au) perpendicular to a molecular outflow traced by the CO emission.  
\newline\par\noindent
\cite{Dunham2006} reported the first Spitzer observations  of IRAM04191 using the InfraRed Array Camera (IRAC). They use these observations and radiative transfer models to show that IRAM04191 has an internal luminosity (i.e. excluding any luminosity arising from heating caused by an external radiation field) of \textit{L$_\mathrm{int}$} = 0.08 $\pm$ 0.04 \Lsun. This is consistent with a Very Low Luminosity Object (VeLLO) and as such, this source is classified as a VeLLO \citep{diFrancesco2007,Vorobyov2017}. The study of VeLLO driven outflows has been of particular interest in recent years \citep{Kim2019} as a means of identifying sub-stellar mass candidates and studying their formation at the earliest stages \citep{Santamaria2021}.
\newline\par\noindent
However, \cite{Dunham2006} also show that when placing the outflow parameters (e.g. bolometric luminosity, outflow mass and outflow force) derived in \cite{Andre1999} and \cite{CFLee2002}~ on the plots relating these quantities (e.g. Fig. 5 and 7 in \cite{Wu2004}) they find that a source with an internal luminosity of \textit{L$_\mathrm{int}$} $\geq$ 1 \Lsun~ is needed to drive the large CO outflow driven by IRAM04191. This is inconsistent with the internal luminosity derived via their radiative transfer modelling.  To explain this inconsistency, they hypothesise that their Spitzer/IRAC observations are a snapshot of IRAM04191 in a quiescent state and that this source may be subject to periods of episodic accretion. 
\newline\par\noindent
The possible episodic mass accretion event in IRAM04191 was further explored in \cite{Anderl2020}~as part of the CALYPSO (Continuum And Lines in Young ProtoStellar Objects) survey using high angular resolution Plateau de Bure Interferometer (PdBI) observations of N$_{2}$H$^{+}$ and C$^{18}$O. Time dependant chemical and radiative transfer models were used to reproduce the line profiles extracted through the emission peaks observed in the integrated intensity maps of N$_{2}$H$^{+}$ and C$^{18}$O (see their Fig. 1). Their results indicate that these line profiles cannot be reproduced using a constant luminosity model based on the current value of internal luminosity for IRAM04191 (0.08 $\pm$ 0.04 \Lsun). Instead, the peak in the N$_{2}$H$^{+}$ emission at 10$\arcsec$ from the source may be caused by a past accretion burst event producing a luminosity 150 times greater than the present day value. Their models predict that if such accretion bursts occur for $\sim$ 1\% of the Class 0 lifetime of IRAM04191 \citep[assumed to be 0.4 - 0.5 $\times$ 10$^{6}$ years; ][]{Evans2009} then, the mass of IRAM04191 could reach 0.2 - 0.25~\Msun~by the end of the Class 0 stage.
\newline\par\noindent
More recently, the work of \cite{Podio2021} used high angular resolution (0$\farcs$5 to 1$\arcsec$) CALYPSO PdBI observations of IRAM04191 in CO~(2--1), SO~6$_{5}$--5$_{4}$, and SiO~(5--4). Their maps extend up to $\sim$ 4$\arcsec$ compared to the previous $\sim$ 2 arcminute maps taken in older single dish observations \citep[e.g.][]{Andre1999,CFLee2002}. A redshifted outflow is detected \textbf{only in CO~(2--1)} from V$_{lsr}$ = 8.2 km.s$^{-1}$ to 21.2 km.s$^{-1}$, extending $\sim$ 4$\arcsec$ to the north-east at a PA of 20$^{\circ}$. Interestingly, they have no detection of the blueshifted component of the jet. 
\newline\par\noindent
The most recent study of IRAM04191 by \cite{Huelamo2026} used a combination of ALMA sub-mm and Very Large Array (VLA) centimetre observations to find that IRAM04191 is a proto-BD binary candidate object with east and west components separated by $\sim$ 80~mas  (11 au at the assumed distance of 140pc). Furthermore, the authors use archival C$^{18}$O (2--1) observations to confirm the presence of a circumbinary disc in Keplerian rotation around the IRAM04191 system. Through fitting of a Keplerian curve to the position-velocity diagram extracted along the disc major axis, the authors derive a dynamical mass of the system of 50 $\pm$ 40 M$_{Jup}$, consistent with a substellar binary candidate.  
\newline\par\noindent
In summary, the literature surrounding IRAM04191 has revealed a vast amount of information about this system. It is a potential binary candidate surrounded by a large, differentially rotating envelope, with hints of episodic accretion. There is also evidence of a large 0.1 pc outflow in CO. The combination of an outflow and lots of surrounding material available for accretion onto IRAM04191 suggests that it may be able to gather enough material to evolve into a stellar system by the end of the Class 0 stage \citep{Anderl2020}. This unique set of features (i.e. large outflow, proto-BD binary candidate, potential episodic accretion etc) makes IRAM04191 a valuable laboratory to study BD and star formation. Hence, the need for continued dedicated studies of this system.   
\newline\par\noindent
The aim of this article is to provide a detailed study of the molecular gas emission in close proximity to IRAM04191 at an angular resolution better than what is available in the current literature for previous molecular gas observations. A dedicated analysis of the molecular emission lines tracing outflowing and infalling gas in close proximity to this source may hold critical evidence needed to study its formation pathway. As such, we focus our study on the morphology and kinematics of three molecular emission lines $^{13}$CO~(3--2), C$^{18}$O~(2--1) and SO~6$_{5}$--5$_{4}$ (Table \ref{line_table}), which are known to trace outflowing and/or infalling gas in young protostellar systems~\citep{Podio2021,Hales2024,Tychoniec2021,Garufi2022}
\newline\par\noindent
This article is organised as follows: Section \ref{Sec2} describes the observations and the steps used to reduce and analyse these data. In section \ref{Sec3} we present our main results, including the morphological and kinematic properties of the reported emission lines. Section \ref{Sec4} provides a discussion of the results. Lastly, in section \ref{Sec5} we summarise our findings and draw our conclusions from them.  

\begin{table}[]
    \centering
    \caption{Properties of IRAM 04191+1522 relevant to this study}
    \begin{tabular}{lcl} \hline\hline
         Property & Value & Reference  \\
         \hline
        Source mass (\Msun) & 0.05$\pm$ 0.04 & H26~$^{a}$\\
        Env. Mass (\Msun) & 0.5  & M19 \\
        Age (yrs.) & 0.8 - 2 ($\times$ 10$^{4}$)  & A99 \\
        $\dot{M}_\mathrm{acc}$ (M$_\odot$.yr$^{-1}$) & 2 $\pm$ 1~($\times$ 10$^{-6}$) & K19  \\
        Disc incl. ($^\circ$) & 64 $\pm$ 7 & H26 \\ 
        Distance (pc) & 140 & A99  \\
        \textit{V$_\mathrm{sys}$} (km.s$^{-1}$) & 6.5 $\pm$ 0.221 & this work~$^{b}$ \\
        \textit{L$_\mathrm{int}$} (\Lsun) & 0.08 $\pm$ 0.04 & D06 \\
         \hline
    \end{tabular}
    \tablefoot{References: A99 = \citet{Andre1999}, D06 = \citet{Dunham2006}, H26 = \citet{Huelamo2026}, K19 = \citet{Kim2019}, M19 = \citet{Maury2019}. \\
    
    $^{a}$ = The measurment of the mass of the system from \cite{Huelamo2026} is a dynamical mass measurment. 
    $^{b}$ = The systemic velocity was estimated from the line spectra presented in this work. The uncertainty is taken as the spectral resolution of the $^{13}$CO~(3--2) cube as it has the lowest spectral resolution of the ALMA data we present (see Table \ref{beam_params_table}). }
    \label{properties table}
\end{table}

\section{Observations \& Data Reduction}\label{Sec2}
IRAM04191 was observed in the millimetre and sub-millimetre range (Band 6 and 7) with the Atacama Large Mm/sub-mm Array (ALMA) in 2016 and in 2019 (Table \ref{alma_obslog}). The molecular line emission data presented here has been obtained from the ALMA science archive from two different observing programmes. The C$^{18}$O~ (2--1) and SO 6$_{5}$--5$_{4}$ emission lines were targeted in 2016 (ID: 2016.1.01284.S, PI: A. Maury) and the $^{13}$CO~(3--2) line was targeted in 2019 (ID: 2019.1.00847.S, PI: P. Sheehan). The calibrated measurement sets (CMs) for these data were obtained from the EU ALMA Regional Centre (ARC) through the European ARC calibrated data service. The CMs for 2016 the 2019 data were generated using version 2024.1.0.8 and version 42866M (Pipeline-CASA56-P1-B) of the ALMA pipeline respectively. Following this, the Common Astronomy Software Applications (CASA, \cite{McMullin2007}, version 6.5.5) package was used to produce the image cubes for each emission line via the \texttt{tclean} task. A Briggs weighting scheme and robust factors of 0.5, 1.0 and 2.0 were chosen for imaging the C$^{18}$O~ (2--1), SO 6$_{5}$--5$_{4}$ and $^{13}$CO~(3--2) lines respectively. These robust factors were selected in order to recover as much flux as possible from the gas surrounding the central source while preserving the spatial resolution required to achieve the scientific objectives. The image cubes were not primary beam corrected. A summary of the final image cube parameters is provided in Table \ref{beam_params_table}. \\

\noindent After the production of the $^{13}$CO~(3--2) line emission image cube, it was noted that the phase centre (RA = 4$^h$21$^m$56.40$^s$, Dec = 15$^{\circ}$29$^m$48.90$^s$ in J2000) of the image cube was not centred at the position of the central source. This was corrected using the \texttt{phaseshift} task in CASA, to the position of IRAM04191 (RA = 4$^h$21$^m$56.90$^s$, Dec =  15$^{\circ}$29$^m$46.05$^s$ in J2000) so, that it matched the phase centres of the C$^{18}$O~ (2--1) and SO 6$_{5}$--5$_{4}$ image cubes. The channel maps, moment maps and position-velocity (PV) diagrams presented in the results sections were produced from these final emission line cubes using a combination of tools available in the Cube Analysis and Rendering Tool for Astronomy (CARTA, version 4.1) \citep{Comrie2021} and custom \textbf{\textit{Python}} scripts. The channel maps for each of the three molecules are included in Appendix \ref{appendix_channel_maps}. 
\newline\par\noindent
We have complemented the ALMA data with archival Spitzer/IRAC observations previously analysed by \cite{Dunham2006}. The details of the data reduction, analysis and interpretation of these data are presented in \citet{Dunham2006}. We include these data to support the interpretation and discussion of the ALMA observations presented in this study.

\noindent\begin{figure*}[h!] 
\centering
  \includegraphics[width = 0.9\linewidth, keepaspectratio = True]{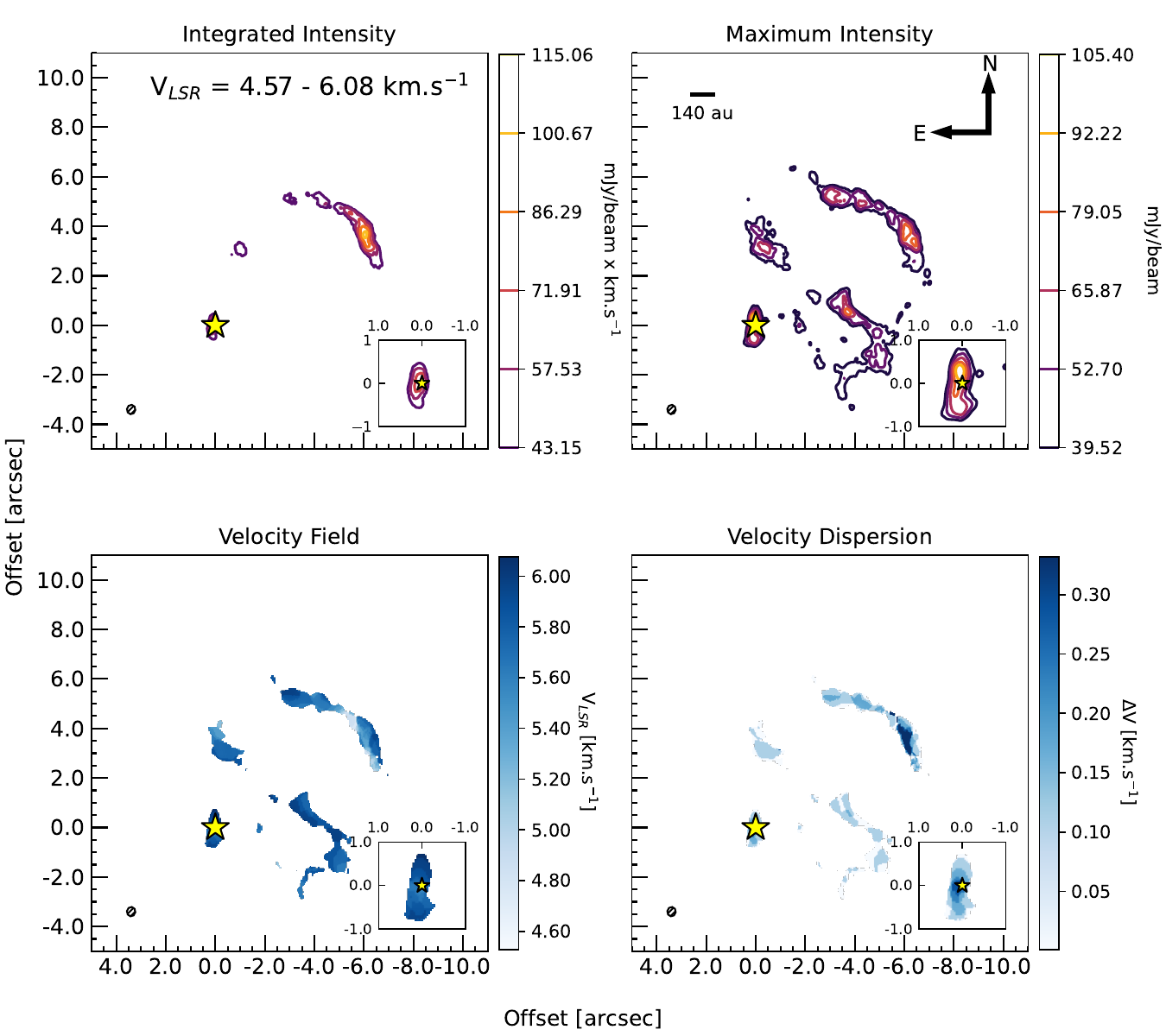}
\captionof{figure}{$^{13}$CO (3--2) blueshifted emission moment maps integrated from 4.57 to 6.08 km.s$^{-1}$. \textbf{Top left:} Integrated intensity map. Contours used were 3 to 8$\sigma$ in steps of 1$\sigma$ ($\sigma$ = 14.38 mJy/beam $\times$ km.s$^{-1}$). \textbf{Top right:} Maximum intensity map. Contours used were 3 to 8$\sigma$ in steps of 1$\sigma$ ($\sigma$ = 13.2 mJy/beam). \textbf{Bottom left:} Velocity field map showing emission from 3$\sigma$ of the integrated intensity map and above (i.e. from 43.15 mJy beam$^{-1}$ $\times$ km.s$^{-1}$ and above). \textbf{Bottom right:} Velocity dispersion map showing emission from 3$\sigma$ of the integrated intensity map and above (i.e. from 43.15 mJy beam$^{-1}$ $\times$ km.s$^{-1}$ and above). In each map there is an small inset plot showing the innermost 2$\arcsec$ centred on the source. The source position is denoted by a yellow star. The beam is represented as a black ellipse in the bottom left corner of each map (0$\farcs$39 $\times$ 0$\farcs$33 at 155.6$^{\circ}$). A spatial scale bar is shown in the top left of the maximum intensity map. The systemic velocity is 6.5 km.s$^{-1}$ (Table \ref{properties table}).}
\label{13CO_blue_moment_maps} 
\end{figure*}

\noindent\begin{figure*}[h!]
\centering
  \includegraphics[width = 0.9\linewidth, keepaspectratio = True]{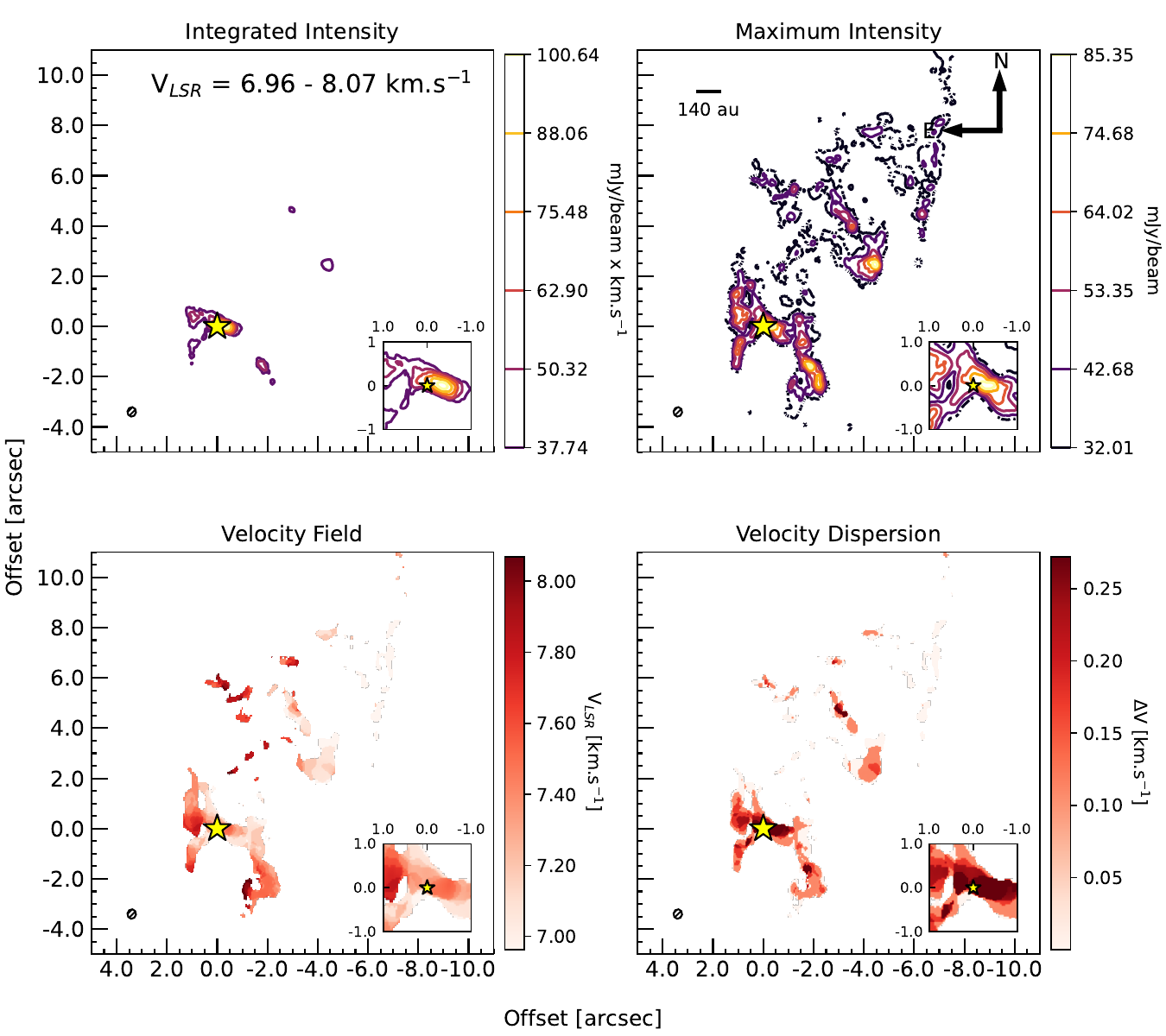}
\captionof{figure}{$^{13}$CO (3--2) redshifted emission moment maps integrated from 6.96 to 8.07 km.s$^{-1}$. \textbf{Top left:} Integrated intensity map. Contours used were 3 to 8$\sigma$ in steps of 1$\sigma$ ($\sigma$ = 12.58 mJy/beam $\times$ km.s$^{-1}$). \textbf{Top right:} Maximum intensity map. Contours used were 3 to 8$\sigma$ in steps of 1$\sigma$ ($\sigma$ = 10.7 mJy/beam). \textbf{Bottom left:} Velocity field map showing emission from 3$\sigma$ of the integrated intensity map and above (i.e. from 37.74 mJy beam$^{-1}$ $\times$ km.s$^{-1}$ and above). \textbf{Bottom right:} Velocity dispersion map showing emission from 3$\sigma$ of the integrated intensity map and above (i.e. from 37.74 mJy beam$^{-1}$ $\times$ km.s$^{-1}$ and above). In each map there is an small inset plot showing the innermost 2$\arcsec$ centred on the source. The source position is denoted by a yellow star. The beam is represented as a black ellipse in the bottom left corner of each map (0$\farcs$39 $\times$ 0$\farcs$33 at 155.6$^{\circ}$). A spatial scale bar is shown in the top left of the maximum intensity map. The systemic velocity is 6.5 km.s$^{-1}$ (Table \ref{properties table}).}
 \label{13CO_red_moment_maps} 
\end{figure*}

\begin{figure*}
    \centering
    \includegraphics[width = 1.05\linewidth, keepaspectratio = True]{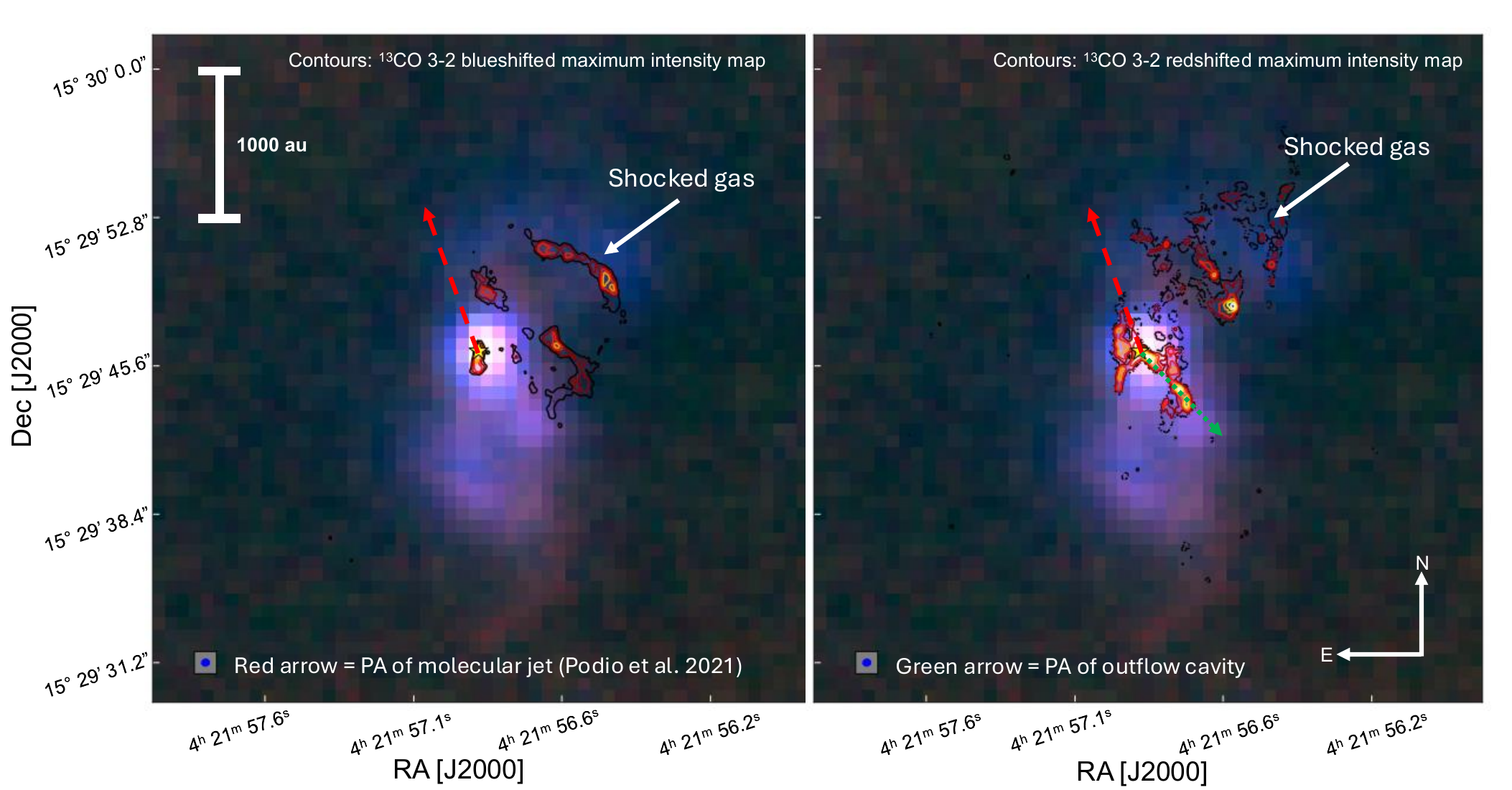}
    \vspace{-0.4cm}
    \caption{\textbf{Left:} Spitzer/IRAC RGB image consisting of the 4.5 (blue), 5.8 (green) and 8$\mu$m (red)  wideband filters compared to the blueshifted maximum intensity emission map of $^{13}$CO~(3--2), presented in Fig. \ref{13CO_blue_moment_maps}. \textbf{Right:} The same IRAC RGB image compared to the redshifted maximum intensity emission map of $^{13}$CO~(3--2), presented in Fig. \ref{13CO_red_moment_maps}. The beam size is shown by the blue ellipse in the bottom left corner of each map. The yellow star marks the source position. The red dashed arrow represents the position angle (20$^{\circ}$) of the redshifted molecular jet detected by \citet{Podio2021}. The green dotted arrow corresponds to the estimated position angle of ($\sim$234$^{\circ}$) of the most extended arm of potentially detected redshifted outflow cavity.}
    
    \label{spitzer rgb_vs_ALMA_13CO}
\end{figure*}

\begin{figure}
    \centering
    \includegraphics[width=\linewidth]{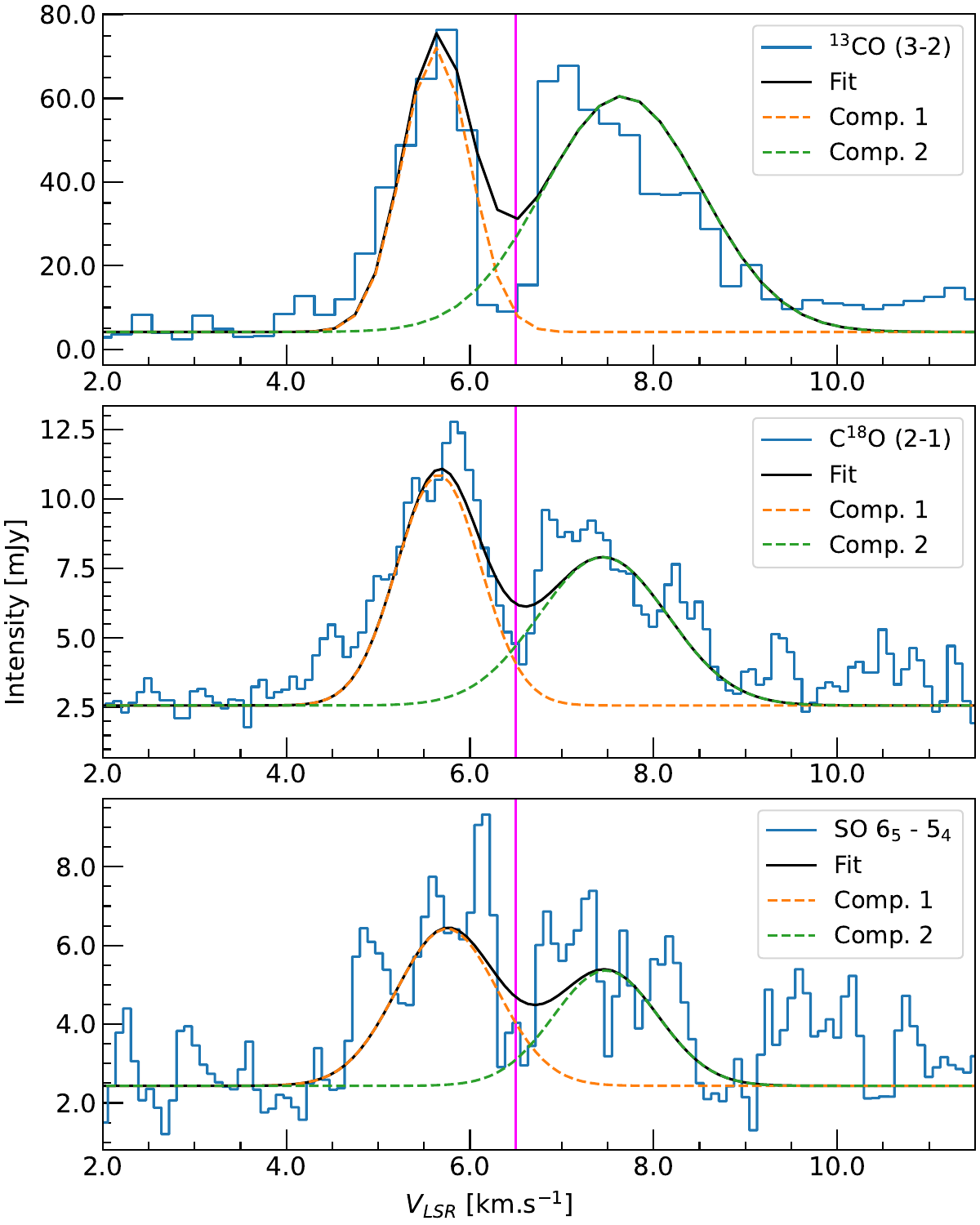}
    \caption{Fitting of the $^{13}$CO~(3-2) line (top),  C$^{18}$O~(2-1) (middle) and SO~6$_{5}$ -- 5$_{4}$ (bottom) lines done using CARTA before continuum subtraction. The spectra were extracted from a circular aperture of radius 0$\farcs$2 centred at the source position. Each Gaussian component in the fit of each line is labelled as either 1 or 2. The magenta line corresponds to the estimated systemic velocity (6.5 km.s$^{-1}$, Table \ref{properties table}). The fitting results are presented in Table \ref{line_fitting_table}.}
    \label{line_fitting}
\end{figure}

\begin{table*}
    \centering
    \caption{Fitting results for the emission lines of interest done in CARTA. The uncertainty range on each measurement are in parentheses. }
    \begin{tabular}{cccccc} \hline\hline
         Line & Comp. &  V$_{c}$  & Intensity & FWHM & Intgr. Flux   \\
         & & [km.s$^{-1}$] & [mJy] &  [km.s$^{-1}$] & [mJy $\times$ km.s$^{-1}$] \\
         \hline 
        
          $^{13}$CO (3-2) & 1 & 5.63 (0.05) & 72.03 (6.71) & 0.87 (0.11) & 63.19 (8.05) \\
           & 2 &  7.66 (0.09) & 60.54 (4.44) & 2.02 (0.23) & 121.04 (11.05) \\ \\
          C$^{18}$O~(2-1) & 1 & 5.66 (0.04) & 10.86 (0.42) & 1.07 (0.08) & 9.44 (0.73) \\
           & 2 & 7.45 (0.07) & 7.90 (0.34) & 1.66 (0.18) & 9.43 (0.89) \\ \\
          SO~6$_{5}$ -- 5$_{4}$ & 1 & 5.75 (0.11) & 6.43 (0.45) & 1.30 (0.25) & 5.55 (1.03) \\
           & 2 & 7.48 (0.15) & 5.37 (0.45) & 1.34 (0.35) & 4.16 (1.04) \\ 
           
           \hline

    \end{tabular}
    \label{line_fitting_table}
\end{table*}

\begin{figure*}[htb]
    \centering
\includegraphics[width = \linewidth]{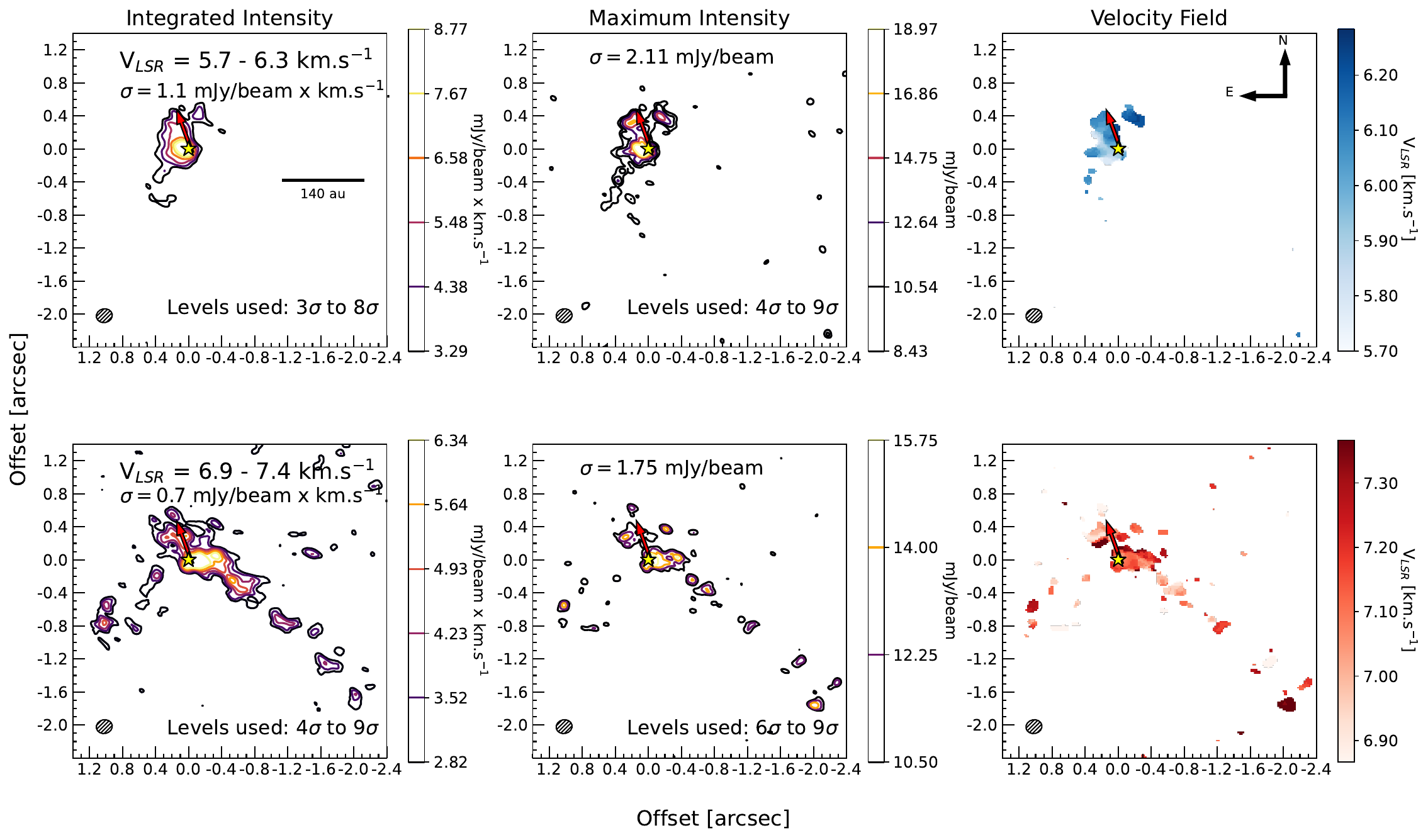}

\caption{\textbf{Left column:} Integrated intensity maps for C$^{18}$O emission line. \textbf{Middle column:} maximum intensity maps for the C$^{18}$O emission line. \textbf{Right column:} Velocity field maps for the C$^{18}$O emission line. The velocity channels used for each row of maps is shown at the top of each map in the left column. The $\sigma$ values used for each map are shown at the top of each respective map. The $\sigma$ levels used for the contours the maps are in 1$\sigma$ steps and the levels used are shown in the bottom right corner of each map. The beam (0$\farcs$19 $\times$ 0$\farcs$17 at 102.1$^{\circ}$) is represented as a black ellipse in the bottom left corner of every map. The source position is denoted as a yellow star. The red arrow shows the position angle (20$^{\circ}$) of the redshifted outflow previously detected by \cite{Podio2021}. }
\label{C18O_Moment0_8_1_maps}
\end{figure*}

\begin{figure*}[h!]
    \includegraphics[width = \linewidth]{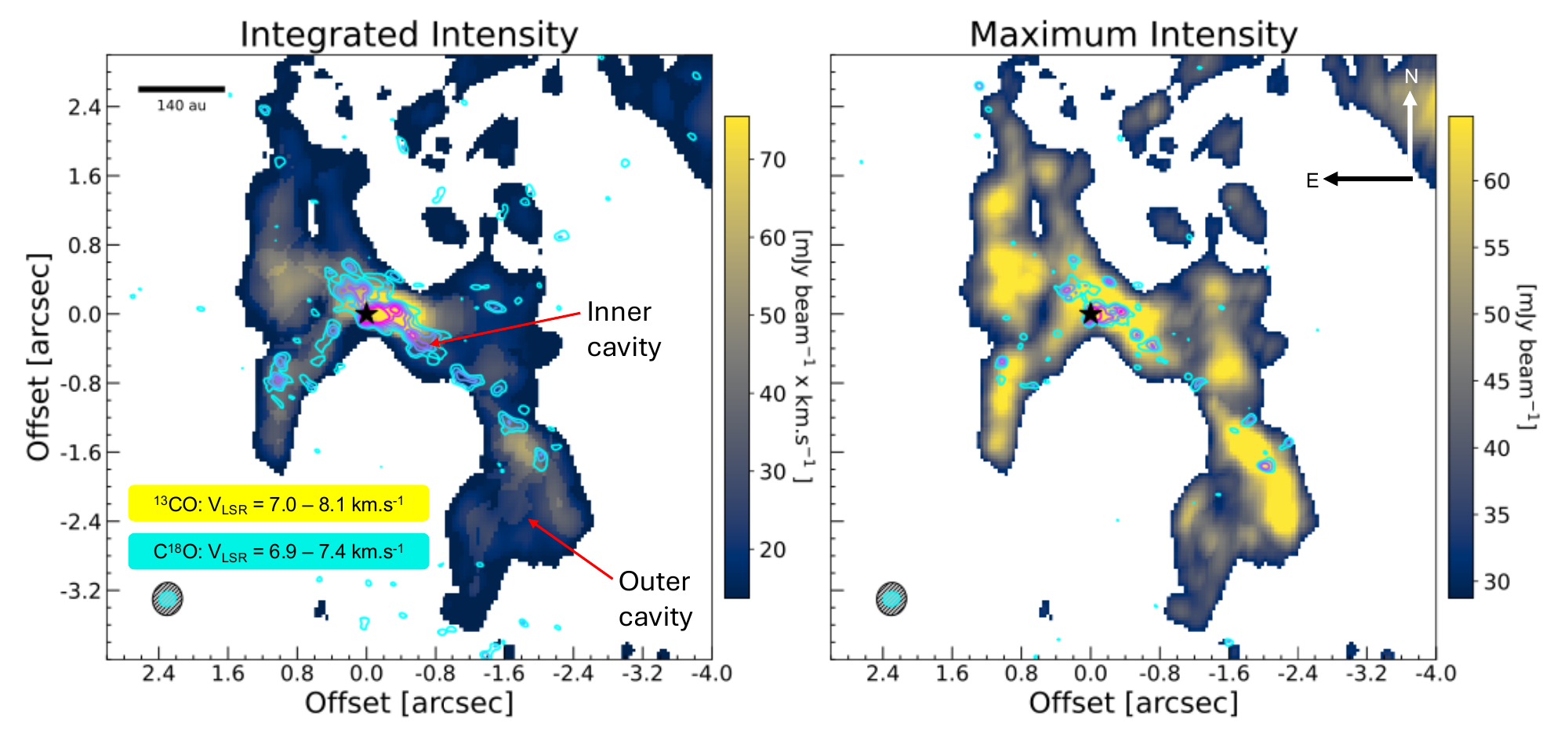}
    \vspace{-0.5cm}
    \caption{Comparison of the redshifted emission of the $^{13}$CO~(3--2) line from V$_{lsr}$ = 7.0 km.s$^{-1}$ to 8.1 km.s$^{-1}$ (represented as a raster map) to the C$^{18}$O~(2--1) line emission integrated from V$_{lsr}$ = 6.9 to 7.4 km.s$^{-1}$ (represented as the the cyan and purple contours). Left = Integrated intensity map. Right = Maximum intensity map. The source position is indicated by a black star. The beams for each map are shown in the bottom left corner, black for $^{13}$CO and cyan for C$^{18}$O (Table \ref{beam_params_table}). The contours used for C$^{18}$O here are the same as the contours used in Fig \ref{C18O_Moment0_8_1_maps} (bottom row). The gas traced by these two molecules at the velocities quoted above show morphology of a similar size and direction. Further analysis (see Appendix \ref{AppendixD}), hint that these emissions are tracing an outflow cavity in this direction.}
    \label{13CO_red_emission_vs_C18O_red_emission}
\end{figure*}

\begin{figure*}[]
    \centering
    \includegraphics[width = 0.95\linewidth]{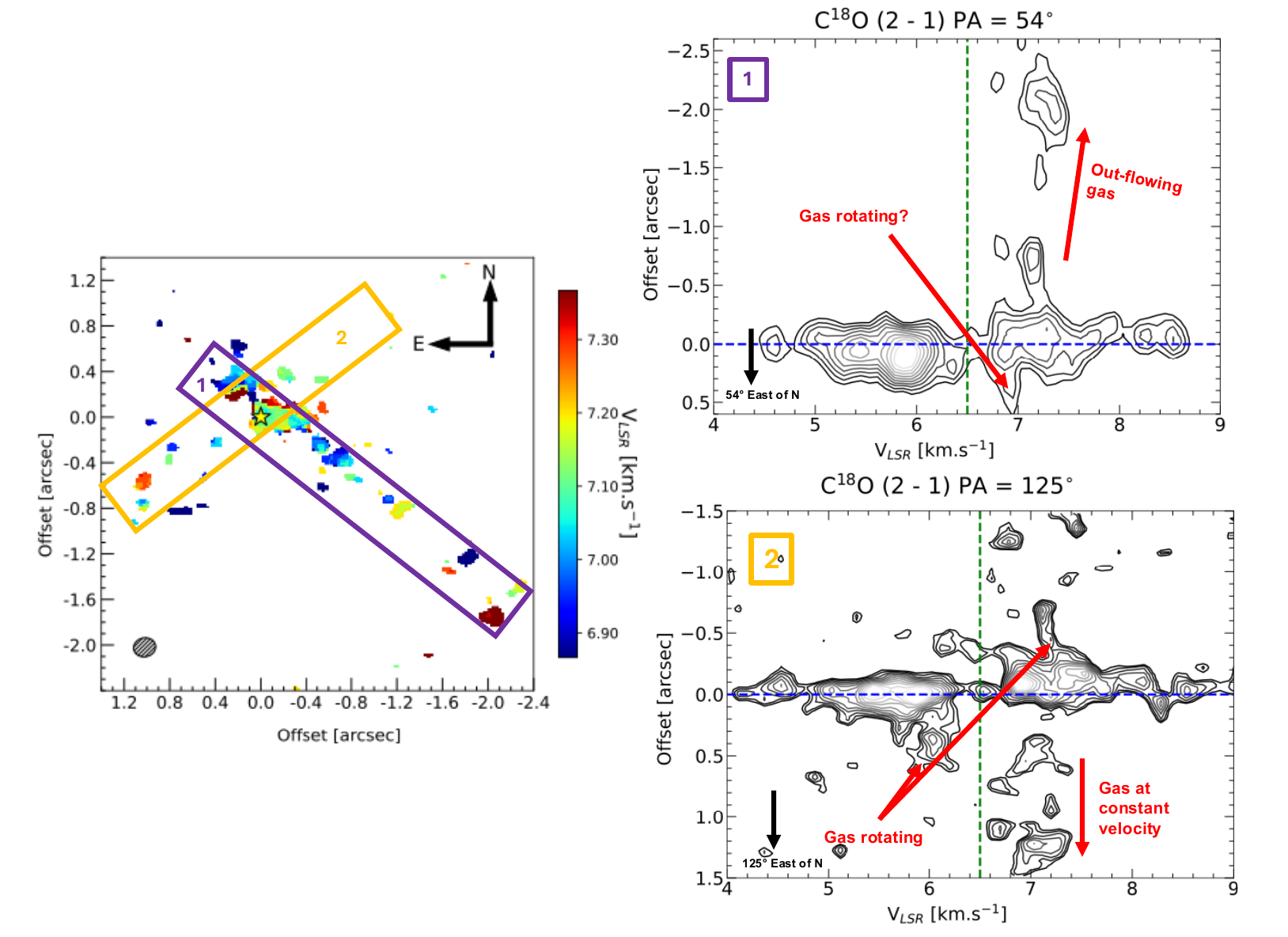}
    \vspace{-0.5cm}
    \caption{\textbf{Left:} The velocity map for C$^{18}$O~(2-1) from V$_{lsr}$ = 6.9 to 7.4 km.s$^{-1}$ (see Fig. \ref{C18O_Moment0_8_1_maps}) with the psudeo-slits overlaid for the corresponding PV diagrams. \textbf{Top right:} PV diagram extracted via slit 1 at PA = 54$^{\circ}$. Slit length = 3$\arcsec$, slit width = 0$\farcs$4 (20 pixels). The contours used were 4.7, 5.8, 7.0, 8.1, 9.3, 10.4, 11.5, 12.7, 13.8, 14.9 and 16.1 mJy/beam. \textbf{Bottom right:} PV diagram extracted via slit 2 at PA = 125$^{\circ}$. The slit was centred at the source position. Slit length= 3$\arcsec$, slit width= 0$\farcs$4 (20 pixels). The contours used were 4.3, 5.7, 7.0, 8.3, 9.7, 11.0, 12.3, 13.7, 13.8, 15.0, 16.4, 17.7 and 19.0 mJy/beam. A Gaussian smoothing filter has been applied to this PV diagram to increase the signal-to-noise. The green and blue dashed lines correspond to the systemic velocity (6.5 km.s$^{-1}$) and zero offset from the source position respectively. The red arrows denote key features in each PV diagram.}

    \label{C18O_pvdiagrams}
\end{figure*}

\begin{figure*}[tbh]
    \centering
\includegraphics[width = 1.05\linewidth]{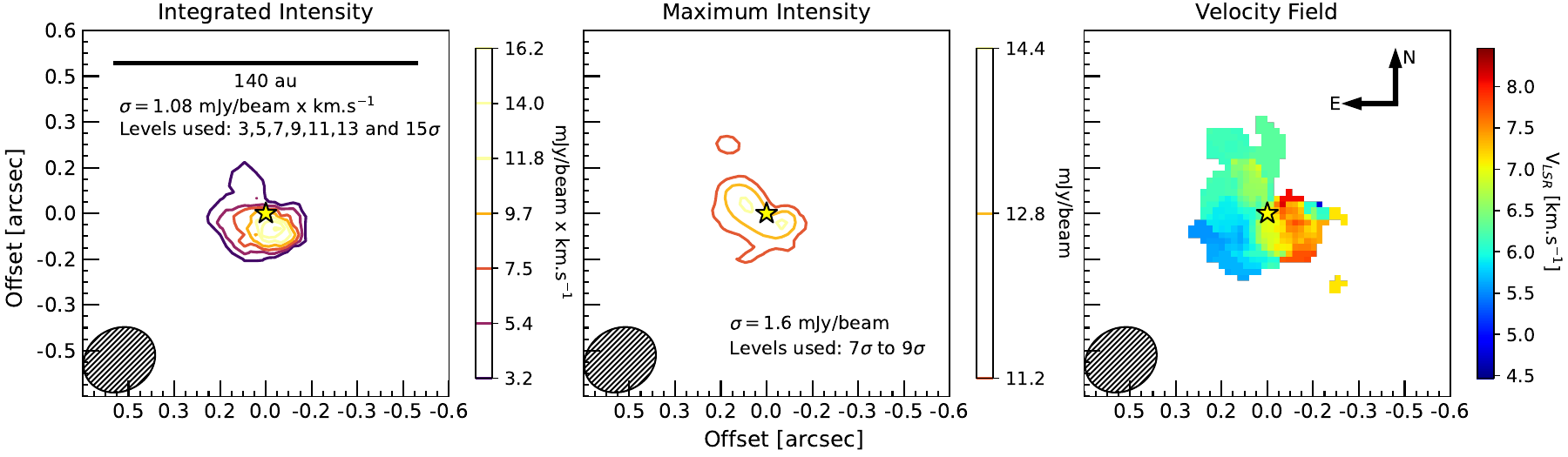}

\caption{SO~6$_{5}$ -- 5$_{4}$ emission line moment maps. \textbf{Left:} Integrated intensity map from 4.5 to 8.5 km.s$^{-1}$. Contours start at 3$\sigma$ ($\sigma$ = 1.08 mJy/beam $\times$ km.s$^{-1}$ and increase in steps of 2$\sigma$. \textbf{Middle:} Maximum intensity map from 4.5 to 8.5 km.s$^{-1}$. Contours start at 7$\sigma$ ($\sigma$ = 1.60 mJy/beam) and increase in steps of 1$\sigma$. \textbf{Right:} Velocity field map. The beam (0$\farcs$24 $\times$ 0$\farcs$20 at PA = 119.4$^{\circ}$) is represented as a black ellipse in the bottom left of each map. The source position is denoted by a yellow star.}
\label{SO_moment_maps}
\end{figure*}

\begin{figure*}[h!]
    \centering
    \includegraphics[width = 0.91\linewidth]{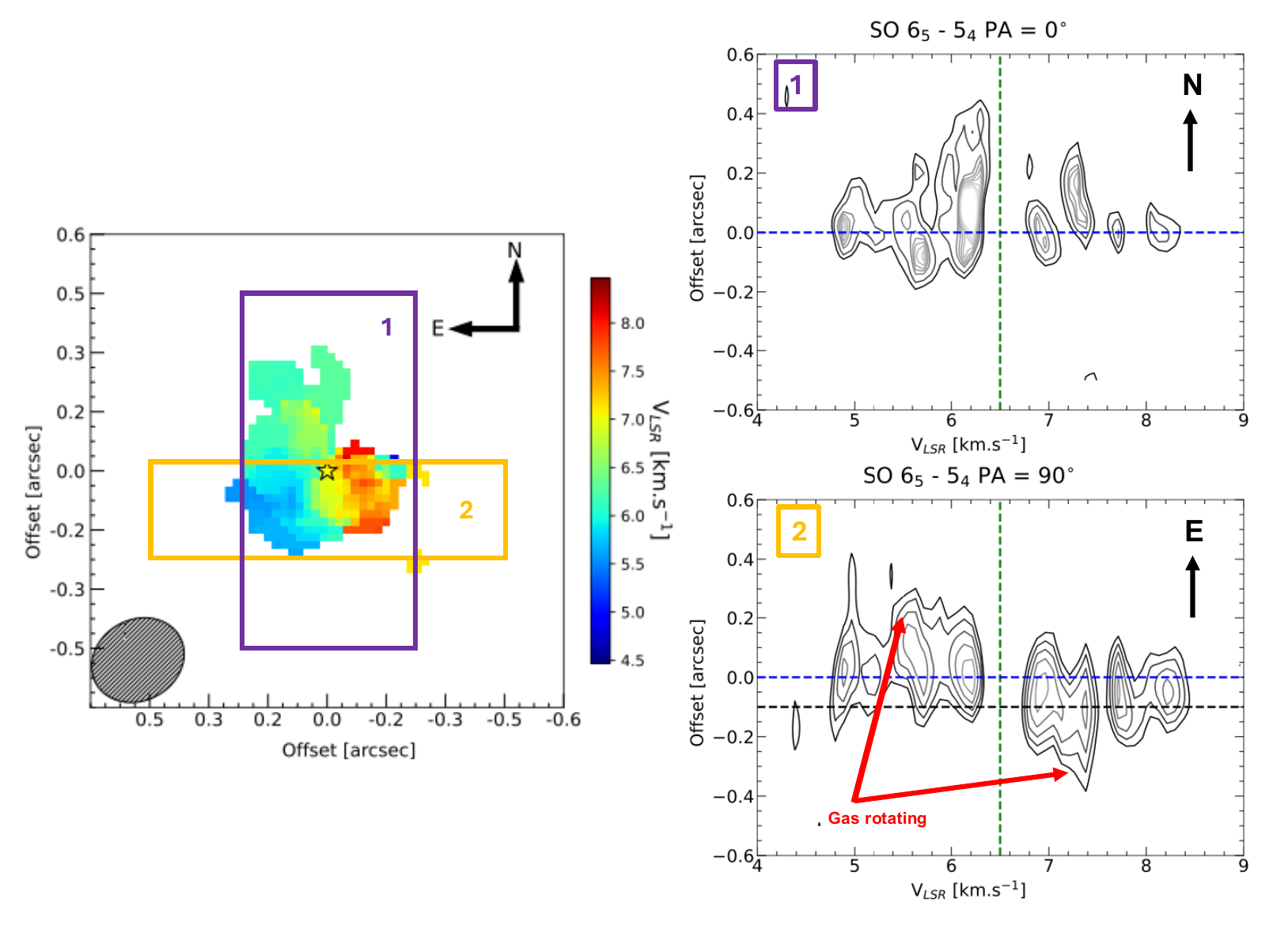}
    43.15\vspace{-0.5cm}
    \caption{\textbf{Left:} The velocity map for the SO~6$_{5}$ -- 5$_{4}$ emission line (see Fig \ref{SO_moment_maps} right panel).  \textbf{Top right:} PV diagram extracted via slit one at PA = 0$^{\circ}$. The slit was centred at the source position. Slit length= 1$\arcsec$, slit width= 0$\farcs$42 (21 pixels). The contours used were 6.5, 7.9, 9.4, 10.0, 10.4, 10.8, 11.2, 11.5, 11.8, 12.0, 12.3 and 12.5 mJy/beam. \textbf{Bottom right:} PV diagram extracted via slit two at PA = 90$^{\circ}$. The slit was centred 0$\farcs$1 below the source position to encapsulate as much of the emission along this PA as possible. Slit length= 1$\arcsec$, slit width= 0$\farcs$24 (12 pixels). The contours used were 5.1, 6.3, 7.5, 8.7, 9.9, 11.0, 12.2 and 13.4 mJy/beam. The black dashed line corresponds to the centre position of the slit. The green and blue dashed lines correspond to the systemic velocity (6.5 km.s$^{-1}$) and zero offset from the source position respectively. }
    \label{SO_band6_pvdiagrams}
\end{figure*}

\vspace{-1cm}
\section{Results}\label{Sec3}
To preface the results section, it is noted here that the systemic velocity adopted for this article is 6.5 km.s$^{-1}$ (Table \ref{properties table}). This was estimated from the spectra of each emission line which are presented later in the article. Also, all the velocities and distances quoted in this article are \textbf{projected}. This projection effect is due the inclination of the system with respect to the plane of the sky, and is estimated to be between 50 - 60$^{\circ}$ \citep[e.g. ][]{Andre1999}. As a consequence, the reported velocities and velocity gradients are \textbf{minimised} by the effect of this projection. 
\subsection{$^{13}$CO~(3--2) Morphology}\label{Sec3.1}
In Fig. \ref{13CO_blue_moment_maps} (top row) we present the integrated and maximum intensity moment maps for the blueshifted emission traced by $^{13}$CO~(3--2). In these maps we observe two main structures traced by the blueshifted $^{13}$CO~(3--2) gas. The first is a large ring shaped filament like structure at a separation of $\sim$ 6$\arcsec$ (measured from the 3$\sigma$ contour of the maximum intensity map) to the north-west. The second is smaller oblong shaped structure centred at the position of the source that extends at most 1$\arcsec$ along the north-south direction. 
\newline\par\noindent
Figure \ref{13CO_red_moment_maps} (top row) shows the integrated and maximum intensity maps for the redshifted $^{13}$CO~(3--2) gas. The morphology of this emission is strikingly different compared to the blueshifted gas showing several structures:
\begin{enumerate}
    \item Sporadic emission is seen from $\sim$ 2$\arcsec$ away from the source to around 10$\arcsec$ in the north-west. This was measured from the 3$\sigma$ contour of the maximum intensity map of this emission. The integrated intensity map shows a much more compact morphology compared to the maximum intensity map, with emission at a separation up to $\sim$ 2$\arcsec$.
    \item  Closer to the source, in the maximum intensity map, we observe a straightened structure to the east of the source that extends along the north-south direction from $\sim$ -2$\arcsec$ to 2$\arcsec$, measured in the same way. 
    \item To the west of the source the redshifted emission begins to curve downwards to around -3$\farcs$5, measured in the same way as outlined above. 
\end{enumerate}
The structures observed to the east and west of IRAM04191 in the redshifted $^{13}$CO~(3--2) gas are clearly connected and may be part of the same structure. We refer to this now as the south-western structure, which is not connected with the sporadic $^{13}$CO~(3--2) redshifted emission seen in the north-west.  
\newline\par\noindent
In Fig. \ref{spitzer rgb_vs_ALMA_13CO} we compare the blue and redshifted maximum intensity maps of $^{13}$CO~(3--2) to the Spitzer/IRAC observations originally presented and discussed in \cite{Dunham2006}. In the left panel, the blueshifted ring shaped structure traced by $^{13}$CO~(3--2) appears to align well with the curved structure seen in the north-west of the IRAC image. This curved structure seems to be primarily traced by the 4.5$\mu$m filter. As noted in \cite{Dunham2006}, the 4.5$\mu$m wideband contains three of the brightest H$_{2}$ emission lines (i.e. 0–0 S(11) at 4.18$\mu$m, 0–0 S(10) at 4.40$\mu$m and 0–0 S(9) at 4.18$\mu$m) which are known tracers of shocked material \citep{Noriega-Crespo_2014}. This spatial overlap suggests that $^{13}$CO~(3--2) blueshifted ring shaped structure could be tracing shocked gas, as seen before in giant molecular clouds and filamentary structures \citep{Zhou2023}. 
\newline\par\noindent
In the right panel of Fig. \ref{spitzer rgb_vs_ALMA_13CO}, we observe that the redshifted spurious $^{13}$CO~(3--2) emission is spread all over the curved structure in the north-west traced by the 4.5$\mu$m emission. The south-eastern structure in the red $^{13}$CO~(3--2) gas seems to tentatively overlap with the base of some extended emission in the south/south-west traced by the 4.5 and 8$\mu$m filters in the IRAC image. Also, as outlined in \cite{Dunham2006} (their Fig. 1), the 4.5$\mu$m emission likely arises from an outflow cavity while the 8$\mu$m filter traces more diffuse emission due to polycyclic aromatic hydrocarbons (PAHs). Given this, the overlap between the south-eastern structure in the redshifted $^{13}$CO~(3--2) gas and the south/south-eastern extended emission in the IRAC image suggests that this redshifted $^{13}$CO~(3--2) gas may be tracing the base of an outflow cavity. This is explored further in Appendix \ref{AppendixD}.

\subsection{$^{13}$CO~(3--2) Kinematics}\label{Sec3.2}
Turning back to Fig. \ref{13CO_blue_moment_maps} (bottom row), the velocity field and velocity dispersion maps for the blueshifted $^{13}$CO~(3--2) gas are shown. These maps are rather complex but, we note the following.
\begin{enumerate}
    \item In the velocity field map (Fig. \ref{13CO_blue_moment_maps}, bottom left panel) the ring shaped filamentary structure shows a complex velocity field throughout. The top half of the ring that faces north-west shows an inner edge that is at a lower local standard of rest (LSR) velocity ($\sim$ 4.6 to 5.4 km.s$^{-1}$) compared to the outer edge of the ring ($\sim$ 5.6 to 6.0 km.s$^{-1}$). 
    \item The inset plot at the bottom right of the velocity field map shows the innermost 2$\arcsec$ closest to the source (i.e. the oblong shape). We note that V$_{lsr}$ decreases from $\sim$ 6.0 km.s$^{-1}$ to about 5.6 km.s$^{-1}$ at the source position. 
    \item In the velocity dispersion map (Fig. \ref{13CO_blue_moment_maps}, bottom right panel) the edge of the ring shaped filamentary structure in the north-west has the largest area of large velocity dispersion with values of $\Delta$V $\geq$ 0.30 km.s$^{-1}$. 
    \item The inset plot at the bottom right of the velocity dispersion map shows the innermost 2$\arcsec$ closest to the source (i.e. the oblong shape). We observe a hint of change in the velocity dispersion of the gas. However, it is noted here that this change is at the level of the velocity resolution of the data cube (0.22 km.s$^{-1}$). The velocity dispersion increases from $\Delta$V $\sim$ 0.05 km.s$^{-1}$ at 1$\arcsec$ away from the source to $\Delta$V $\sim$ 0.30 km.s$^{-1}$ at the source position. A 0.25 km.s$^{-1}$ increase. Further observations are needed to confirm if this observed increase is real or an effect due to insufficient spectral resolution. 
\end{enumerate}

\noindent The velocity field and velocity dispersion maps for the redshifted $^{13}$CO~(3--2) emission are shown in Fig. \ref{13CO_red_moment_maps} (bottom row). As for the blueshifted emission, the velocity field and dispersion maps are complex however we note the following.
\begin{enumerate}
    \item In the velocity field map (Fig. \ref{13CO_red_moment_maps}, bottom left panel) the spurious emission to the north-west is mixed in terms of its velocity. The most extended emission to the north west is at $\sim$ V$_{lsr}$ = 7.0 km.s$^{-1}$. This velocity increases as we move closer to the south-eastern structure that surrounds the source. 
    \item In the structure to the south-east, we observe a chaotic velocity field. The velocity field appears fastest at the eastern and south-western edge of this structure, but the field mixes into different velocities as we move closer to the source. 
    \item The inset map, showing the innermost 2$\arcsec$ from the source, at the bottom right of this velocity field map  shows that the velocity is higher to the east of the source ($\sim$ 7.8 km.s$^{-1}$) than to the west of the source ($\sim$ 7.4 km.s$^{-1}$).
    \item In general, the velocity dispersion map for the redshifted $^{13}$CO~(3--2) emission shows the highest dispersion in the innermost 2$\arcsec$ region close to the source.  
\end{enumerate}
Lastly, in Fig. \ref{line_fitting} (top panel) we present a line profile extracted from the $^{13}$CO~(3--2) emission line cube from the innermost 0$\farcs$2 radius from the source. The $^{13}$CO~(3--2) line profile is double peaked and we have fitted it with two components (see Table \ref{line_fitting_table}). The blueshifted component of the fit is narrower and has a higher intensity than the redshifted component. The double peaked shape of this profile may indicate the presence of gas associated to a circumstellar disc in rotation or envelope material in rotation. The presence of an extended red wing may be indicative of an outflow \citep{Ray2021}. However, we suspect that the shape of this line profile may be caused by a mix of different emission components in the $^{13}$CO~(3--2) gas in the innermost region. This is supported by the complex morphology, velocity field and velocity dispersion maps for both the red and blueshifted $^{13}$CO~(3--2) gas in the 2$\arcsec$ region around the central source. 

\subsection{C$^{18}$O~(2--1) Morphology}\label{Sec3.3}
Figure \ref{C18O_Moment0_8_1_maps} shows the integrated intensity (left column), maximum intensity (middle column) and velocity field (right column) maps over two velocity ranges (5.7 to 6.3 km.s$^{-1}$ and 6.9 to 7.4 km.s$^{-1}$). These channels were selected as we detect the most prominent C$^{18}$O~(2--1) emission features in the blue and redshifted side of the C$^{18}$O~(2--1) line respectively. In this section we discuss the results from the integrated and maximum intensity maps over these two regions. 
\newline\par\noindent
Firstly, we discuss the morphology of the 5.7 - 6.3 km.s$^{-1}$ velocity channels. In both the integrated and maximum intensity maps, we find two extended components in the C$^{18}$O~(2--1) gas.  We detect a north-eastern component. We observe it out to $\sim$ 0$\farcs$5 (measured from the 3$\sigma$ contour). We note that this material lies close to the PA (30$^{\circ}$) of the previously detected large scale outflow \citep{Andre1999,CFLee2002,CFLee20052005}. A second component is also observed along the north-west to south-east direction. The north-western part is of $\sim$ 0$\farcs$6 while the south-eastern part of this component is observed out to $\sim$ 0$\farcs$7. The total projected length of this component is $\sim$ 1$\farcs$3. These distances are measured from the 3$\sigma$ contour of the integrated intensity map. 
\newline\par\noindent
The second velocity range (6.9 to 7.4 km.s$^{-1}$) also shows two components, that are the most extended components observed in the C$^{18}$O~(2--1) gas. The first is the south-east component observed in the 5.7 - 6.3 km.s$^{-1}$ range. The PA has changed slightly and this component now extends out to $\sim$ 0$\farcs$8 (measured from the 4$\sigma$ contour). The second and new component is detected in the south-west direction, and displays a narrow knotty emission that extends out to $\sim$ 2$\arcsec$ from the source in both the integrated and maximum intensity maps (again, measured from the 4$\sigma$ contour). These knot like features may appear due to the high angular resolution of the observations (emission more extended than 2$\arcsec$ in this line is filtered-out by the interferometer) and faintness of the emission. 
\newline\par\noindent
From the analysis of the 6.9 to 7.4 km.s$^{-1}$ velocity channels in C$^{18}$O~(2--1) we note a comparable morphology between the C$^{18}$O~(2--1) gas in these channels and the redshifted emission seen in $^{13}$CO~(3--2). In Fig. \ref{13CO_red_emission_vs_C18O_red_emission} we compare these two emissions. A striking similarity in the shape of the gas traced in the south-east by both lines is observed. The C$^{18}$O~(2--1) emission in this velocity range appears to fit inside the redshifted $^{13}$CO~(3--2) emission. As previously mentioned in section \ref{Sec3.1}, the redshifted $^{13}$CO~(3--2) gas may be tracing the base of an outflow cavity (see Appendix \ref{AppendixD} also). The spatial correspondence between the two tracers suggests that the C$^{18}$O (2–1) emission in these channels may originate from a denser component of gas located at the base of the outflow cavity. This is interesting given there is recent evidence of nested outflow morphology of atomic and molecular emission in several young low mass YSOs \citep{Birney2024,Pascucci2025}. However, we do note that this effect may in part be due to the fact that C$^{18}$O~(2--1) traces colder and denser regions than $^{13}$CO~(3--2). 

\subsection{C$^{18}$O~(2--1) Kinematics}\label{Sec3.4} 
Fig. \ref{line_fitting} (middle panel) shows the line profile of C$^{18}$O~(2--1) at the innermost 0$\farcs$2 radius from the source. This profile is quite similar in shape to the line profile of $^{13}$CO~(3--2). It shows two components, with the blue side having a narrower and higher peak compared to the red side. As such, we apply a similar argument to this line profile as for $^{13}$CO~(3--2). The C$^{18}$O~(2--1) line profile may indicate there is gas associated to a disc in rotation. However, we note here that a double peaked C$^{18}$O~(2--1) line profile may be caused by self absorption of warm photons closer to the central source being absorbed by cold gas in the rotating envelope \citep{Areal2018}. As before, the exact shape of the line profile is probably due to mix of different emission features in the gas. The results of our line fitting is presented in Table \ref{line_fitting_table}.
\newline\par\noindent
Now, we discuss the velocity field maps for C$^{18}$O~(2--1) in the two velocity channel regions (Fig. \ref{C18O_Moment0_8_1_maps}, right column). From V$_{lsr}$ = 5.7 to 6.3 km.s$^{-1}$, the gas seems to be slower closer to the source and faster as we move out along the extension to the north-east. However, the slowest gas with respect to the systemic velocity is seen as a blob at $\geq$ 6.3 km.s$^{-1}$ at around 0$\farcs$4 in the north-west. Next, we draw attention to the velocity field map for the channels from V$_{lsr}$ = 6.9 to 7.4 km.s$^{-1}$. We observe knot-like emission features along the south-west and south-east directions at various redshifted velocities. These knot-like emission features are caused by the filtering of certain spatial scales by the ALMA interferometer in this high angular resolution array configuration. We observe a flat extended structure around and to the west of the source of up to $\sim$ 0$\farcs$5. Most of this structure is at a velocity of 7.15 km.s$^{-1}$, but its outer edge seems slightly faster, although the emission may be a superposition of different components.
\newline\par\noindent
Finally, we present two PV diagrams extracted from the C$^{18}$O~(2--1) cube. Two pseudo-slits were placed at PAs of 54$^{\circ}$ and 125$^{\circ}$ on the image cube. The velocity field map from V$_{lsr}$ = 6.9 to 7.4 km.s$^{-1}$ is used for visualisation purposes (Fig \ref{C18O_pvdiagrams}, left panel). Two PV diagrams were extracted, one for each pseudo-slit. In Fig. \ref{C18O_pvdiagrams} (top right panel) is the PV diagram extracted from the first pseudo-slit at 54$^{\circ}$. We note two key features in this diagram. The first is some extended, non continuous emission that starts at $\sim$ - 0$\farcs$8 and is seen out to $\sim$ -2$\farcs$4. The velocity increases from $\sim$ 7.1 km.s$^{-1}$ at - 0$\farcs$8 to $\sim$ 7.4 km.s$^{-1}$ at - 2$\farcs$4. These positions indicate that this PV diagram is tracing emission from the south-western component observed in C$^{18}$O~(2--1) from V$_{lsr}$ = 6.9 to 7.4 km.s$^{-1}$ at an estimated PA of 234$^\circ$. The increase in velocity with distance from the source is referred to as a positive velocity gradient, which is indicative of accelerating material typical of the outflow launching process \citep{Ferreira1997,Pascucci2023}. Hence, this feature of the PV diagram shows that the south-western component is likely associated to out-flowing gas \citep[e.g.][see also Appendix \ref{AppendixD}]{Whelan2014,Riaz2017}. The second feature is a slightly offset bump of 0$\farcs$5 at 7 km.s$^{-1}$ which might be part of gaseous emission from the disc \citep{Huelamo2026}. 
\newline\par\noindent
In Fig. \ref{C18O_pvdiagrams} (bottom right panel) we show the PV diagram extracted from the second pseudo-slit at 125$^{\circ}$. Two features are also observed in this diagram. The first is a feature at a constant velocity of V$_{lsr}$ $\sim$ 7 km.s$^{-1}$ moving away from the source position from 0$\farcs$4 to 1$\farcs$5. This suggests the south-eastern component seen in C$^{18}$O~(2--1) from V$_{lsr}$ = 6.9 to 7.4 km.s$^{-1}$ may be part of an outflow cavity \citep[e.g.][ see Appendix \ref{AppendixD} also]{Stahler1994,Arce2013}. The second feature in this PV diagram is \textbf{Keplerian-like} (marked by the red arrows) pattern. This indicates there may be gas in rotation around the central source. However, we do note there is limited spatial and spectral resolution in our PV diagram, to present a definitive conclusion. The Keplerian motion of the disc around IRAM 04191 has been confirmed using this line before and those results are presented in \cite{Huelamo2026}.
\newline\par\noindent
\subsection{SO~6$_{5}$--5$_{4}$ Morphology}\label{Sec3.5}  
Figure \ref{SO_moment_maps} presents the integrated intensity (left panel) and maximum intensity (middle panel) maps for the SO~6$_{5}$--5$_{4}$ line. The emission traced by this line is much more compact compared to the gas traced by the CO isotopologues presented in the previous sections. The gas traced by SO~6$_{5}$--5$_{4}$ spans $<$ 0$\farcs$4 in any direction. Firstly, an  emission of $\sim$ 0$\farcs$2 along the east-west direction and slightly south of the position of IRAM04191 is observed. A second component is seen to the north of IRAM04191 by about 0$\farcs$25. The maximum intensity map shows what appears to be a mix of these two structures. The peak contours in this map lay just to the west and north-east of IRAM04191. Above the north-east peak contour is a blob of emission about 0$\farcs$28 north-east from the central source.

\subsection{SO~6$_{5}$--5$_{4}$ Kinematics}\label{Sec3.6} 
In Fig. \ref{line_fitting} the line profile for SO 6$_{5}$ -- 5$_{4}$ in the innermost 0$\farcs$2 radius is shown. Like the line profiles of the CO isotopologues, this profile contains a blue and redshifted peak. However, unlike the line profiles of $^{13}$CO~(3--2) and {C$^{18}$O~(2--1) the intensity and FWHM of each component of the SO~6$_{5}$--5$_{4}$ line profile are comparable. This may indicate that extended emission laying along the east-west direction could be related with a gaseous disc that has been previously observed in C$^{18}$O \citep{Huelamo2026}. 
\newline\par\noindent
The right panel of Figure \ref{SO_moment_maps} presents the velocity field map of the SO~6$_{5}$--5$_{4}$ emission. Like the integrated intensity map it shows the northern structure above the central source and the extended emission along the east-west direction. In the velocity field map, we observe blueshifted emission in the east and redshifted emission in the west. This is strong evidence that the gas detected in the east-west direction is associated with rotation. Also, note that this rotation pattern is similar to the one observed in the C$^{18}$O~(2--1) observations that revealed the disc \citep{Huelamo2026}. 
\newline\par\noindent
We have generated two PV diagrams using the SO emission cube. They have been extracted using two pseudo-slits at 0$^{\circ}$ and 90$^{\circ}$ (see Fig \ref{SO_band6_pvdiagrams}, left panel). The PV diagram extracted using the first pseudo-slit is shown in the top right panel of Fig. \ref{SO_band6_pvdiagrams}. The slit was placed along the north-south direction with the aim of capturing the emission from the northern extended structure. The outer contour in this PV diagram which extends to an offset of $\sim$ 0$\farcs$425 is at an V$_{lsr}$ approximately equal to the systemic velocity. Considering that this northern emission is only extended from the central source by $\sim$ 0$\farcs$3, hints that this material may be gravitationally bound. \\

\noindent The PV diagram extracted via the second pseudo-slit is shown in the bottom right panel of Fig. \ref{SO_band6_pvdiagrams}. The slit was placed along the east-west direction. This PV diagram shows that the blueshifted side at 5.4 km.s$^{-1}$ has a positive offset from the central source of $\sim$ 0$\farcs$3 and the redshifted side at 7.4 km.s$^{-1}$ has a negative offset of $\sim$ - 0$\farcs$4. This gradient in velocity close to the central source is consistent with gas  rotating anti-clockwise, as observed previously by \cite{Maury2020}. Give that a gas disc has been detected around IRAM04191 along this direction in C$^{18}$O also rotating counter-clockwise \citep{Huelamo2026}, we suggest that this flattened like structure along the east-west direction seen in SO~6$_{5}$--5$_{4}$ could be part of the disc that has not yet collapsed fully and therefore is not yet in Keplerian rotation. This may be possible given the estimated age of this system. 

\section{Discussion}\label{Sec4}
From our analysis and the literature, IRAM04191 has observational properties that are consistent with a star-like formation scenario. In the literature, at large spatial scales ($\sim$ 2 arcminutes), there is evidence of an outflow \citep{Andre1999}, outflow cavity \citep{CFLee2002,CFLee20052005} and a large rotating natal envelope \citep{Belloche2002}. More recent studies \citep[e.g.][]{Podio2021} have detected the presence of a redshifted molecular jet. The following subsection discusses the findings of this work in the context of the existing literature. 

\subsection{Comparison of morphology and kinematics of our $^{13}$CO emission to existing literature. }

\cite{Andre1999} discovered IRAM04191 via the detection of a large-scale molecular outflow with both blue and redshifted lobes traced in single dish observations of CO (2--1). The blueshifted CO (2--1) lobe was integrated over 0 - 5 km.s$^{-1}$ while the redshifted CO lobe was integrated over 8 - 13 km.s$^{-1}$. Each of the outflow lobes extends $\sim$ 2 arcminutes ($\sim$ 0.1 pc at the distance to IRAM04191; Table \ref{properties table}). Lastly, the authors indicate a systemic inclination of $\sim$ 50 - 60$^{\circ}$, and they do not provide an estimate for the mass contained within the outflow. In contrast, the largest spatial scales probed by the observations presented in this work are $\sim$ 10$\arcsec$ as probed by the redshifted $^{13}$CO (3--2) maximum intensity map. As noted above, the morphology of the $^{13}$CO (3--2) emission traces a mix of potentially shocked gas with a ring like morphology in the blue (integrated from 4.57 to 6.08 km.s$^{-1}$) and a spurious spread morphology in the redshifted gas (integrated from 6.96 to 8.07 km.s$^{-1}$). Additionally, the redshifted $^{13}$CO (3--2) emission indicates the presence of a potential outflow cavity in the south/south-west (see also Appendix \ref{AppendixD}). The other emission lines presented in this article trace much smaller spatial scales than the single dish  work of \cite{Andre1999}. We trace gas in close proximity to IRAM04191. Mass estimates at the spatial scales probed by \cite{Andre1999} are not calculated from the $^{13}$CO (3--2) data presented in this study as our observations are not at a comparable spatial scale or sensitivity to \cite{Andre1999} and suffer from the effects of spatial filtering by the ALMA interferometer. 
\newline\par\noindent
The study of \cite{CFLee20052005}, also present large-scale observations of IRAM04191 in the N$_2$H$^+$ (1-0), HCO$^+$ (1--0), and CS (2--1) emission lines with BIMA. They compared this data with CO (1--0) observations presented by \cite{CFLee2002}. Like \cite{Andre1999}, the large outflow is best detected in CO, however, the centre of the CO(1--0) outflow is observed to be offset from IRAM04191 by 4$\arcsec$ to the north-west. The CO map of \cite{CFLee20052005} is integrated over 25.4 km.s$^{-1}$. In addition, they detect two CO outflow cavities to the south-east. One cavity is 80$\arcsec$ long while the other is more extended but further to the south. The PAs of these cavities are 30 and 18$^{\circ}$. \cite{CFLee20052005} also provide a mass estimate of 0.02~\Msun~for the CO emission assuming optically thin emission, an excitation temperature of 20 K and a CO abundance relative to molecular hydrogen of 8.5$\times$10$^{-5}$.  In comparison, the cavity observed in the $^{13}$CO and C$^{18}$O data presented here is along the south-west direction and is much less extended compared to the cavities observed by \cite{CFLee20052005}. As stated previously, mass estimates from the $^{13}$CO and C$^{18}$O data presented here is inhibited by the effects of spatial filtering by the interferometer. It is clear that the morphology and velocity ranges of the gas traced by these low angular resolution, large spatial scale studies of IRAM04191 vary dramatically from the high spatial resolution small spatial scale observations presented by this paper.   
\newline\par\noindent
However, \cite{Podio2021} provides observations of IRAM04191 that are on a more comparable spatial scale to our observations.  \cite{Podio2021} observe IRAM04191 with BIMA targeting the CO (2--1), SO ($5_{6}$--$4_{5}$) and SiO (5--4) emission lines. They detect a molecular jet only in CO (2--1) (CO beam size: 1.07$\arcsec$ $\times$ 0.83$\arcsec$). They observe a redshifted monopolar molecular jet extended in the north-east direction by $\sim$ 4$\arcsec$ at a terminal velocity of 13 km.s$^{-1}$ with respect to the systemic velocity. The CO emission in \cite{Podio2021} was integrated from 8.2 to 21.2 km.s$^{-1}$. They do not provide a mass estimate for this red molecular jet. In comparison, in our $^{13}$CO data we detect shocked gas in blue and red channels extended by up to 10$\arcsec$ in the north-west. The outflow cavity that we find in both the $^{13}$CO and C$^{18}$O is extended by $\sim$ -2$\farcs$4 to the south-west and is also redshifted.  

\subsection{The fate of IRAM04191}
\noindent Current observations tend to favour BD formation as a scaled-down version of star formation \citep[see][and references therein]{Palau2024}. As such, we would expect to observe shorter, slower and less powerful outflows, lower mass accretion rates etc.  So, we now compare what has been observed for both low mass protostars and young BDs. 
\newline\par\noindent
The formation process of low mass stars in the Class 0/I stage is characterised by both mass accretion and mass-loss processes. Molecular outflows of length 0.1 to 5 pc with velocities in the order of $\sim$ 10 km.s$^{-1}$ have been observed in a variety of tracers. They are interpreted as material swept up by the faster, inner molecular jet \citep{Frank2014,Bally2016}. Episodic accretion bursts have been detected in the Class 0/I stage for stellar masses as low as 0.1~\Msun \citep{Stock2020}. These bursts are generally associated to interactions with disc material. Streamers of in-falling material interacting with the circumstellar disc may play a part in explaining the origin episodic accretion. They have been observed for a handful YSOs at a variety of spatial scales (both in and outside the natal envelope) at the Class 0/I stage \citep{Pineda2020,Hales2024,Valdivia-Mena2024}.
\newline\par\noindent
In comparison, for BDs, \cite{Whelan2005,Riaz2017} report the first atomic outflows launched by a BD and proto-BD respectively. \cite{Whelan2005} reports a projected length and blueshifted velocity of 0$\farcs$08 to 0$\farcs$1 (10 to 12.5 au at the assumed distance of 125 pc) at approximately - 40 km.s$^{-1}$ for the outflow launched by the BD $\rho$-Ophiuchi. Whereas, \cite{Riaz2017} reports a projected length of 0.26 pc ($\sim$ 5.36 $\times 10^{4}$ au at an assumed distance of 387 $\pm$ 1.3 pc) and blueshifted velocity of $\sim$ - 45 km.s$^{-1}$ for the outflow driven by the proto-BD M1701117. In terms of accretion processes, with particular emphasis on streamers, \cite{Riaz2024ObservationsDwarf} report the only known detection of an accretion streamer in the vicinity of a proto-BD candidate to date. The reported projected length of this streamer is 2000 to 3000 au. This is smaller in scale to the streamers seen in the vicinity of low mass protostars (e.g. \cite{Pineda2020}). 
\newline\par\noindent
The northern structure of gas detected in SO~6$_{5}$--5$_{4}$ close to IRAM04191 is of a projected length of 0$\farcs$3 (42 au at the assumed distance to IRAM04191). The map size and spectral resolution of the observations presented here inhibit our ability to determine its origin. If this northern structure is later determined (i.e. with high spectral resolution maps in streamer tracing lines e.g. HCO$^+$ at intermediate spatial scales) to be part of an accretion streamer then it might explain the suggested episodic accretion events in this system. Also, considering the length scale of the known CO outflow detected by \cite{Andre1999}, which seems more typical of low-mass protostars, and the small projected size of this northern structure, IRAM04191 shows observational characteristics of both low-mass YSOs and BDs. Lastly, we note that the flattened structure in SO~6$_{5}$--5$_{4}$ along the east-west direction of IRAM04191 has comparable velocity structure and morphology to the flattened and rotating material detected around this source in C$^{18}$O~(2--1) and the subsequent detection of a Keplerian disc in the same emission line \citep{Huelamo2026}.
\newline\par\noindent
So, what will be the fate of IRAM04191 by the end of the Class 0 stage? Our results present mixed indications. Observations in  SO~6$_{5}$--5$_{4}$ detect the presence of a potentially gravitationally bound structure to the north of IRAM04191. The origin of this structure is unclear and warrants further investigation. In C$^{18}$O (2--1) and $^{13}$CO (3--2) we present observations that appear to be indicative of an outflow cavity to the south/south-west and indications of shocked gas in $^{13}$CO (3--2) to the north-west. Alongside this, previous evidence of large and small scale mass-loss processes (outflows and outflow cavities; \cite{Andre1999,CFLee2002}) add further complication to the mass assembly picture for IRAM04191. Hence, we cannot conclude on the fate of this object as more observations are needed for two reasons.

\begin{enumerate}
    \item Further observations are required at high spectral resolution and at spatial scales in-between the results presented here, and the large scale structures reported in the literature \citep{Andre1999,CFLee2002,CFLee20052005}. These proposed observations will be critical to assess the link between large and small scale mass-loss and mass accretion processes.
    \item A monitoring campaign with observations taken at a high cadence in time are needed to confirm if this source goes into outburst because of episodic accretion events. Other VeLLOs, for example IRAS 15398–3359, have also been suggested to be out-bursting \citep{Jorgensen2013}. If these outbursts are confirmed it will provide strong evidence that IRAM04191 may evolve into a stellar system by the end of the Class 0 stage. These observations may also tell us if these outbursts are EXor like or FUor like \citep{Fischer2023}.
\end{enumerate}

\section{Summary \& Conclusions}\label{Sec5}
We have carried out a detailed morphological and kinematic study of high density gas in the immediate proximity of IRAM04191 as traced by three emission lines $^{13}$CO~(3--2), C$^{18}$O~(2--1) and SO~6$_{5}$--5$_{4}$. Our summary and conclusions are as follows: 
\begin{enumerate}

    \item Emission from $^{13}$CO~(3--2) is complex. The blueshifted emission traces gas on larger spatial scales compared to C$^{18}$O~(2--1) or SO~6$_{5}$--5$_{4}$. The ring-like structure in the blueshifted gas overlaps with the arc of gas traced by the 4.5$\mu$m passband to the north-west of the Spitzer/IRAC imaging. Hence, this ring-like structure may be tracing previously reported shocked gas. 

    \item The redshifted $^{13}$CO~(3--2) gas is also complex. The spurious emission to the north-west may also be tracing shocked gas. Whereas, the structure to the south/south-west appears to be tracing the base of an outflow cavity. The presence of an outflow cavity is important as it naturally leads to the idea of a secondary outflow driven by the IRAM04191 system in a previously unreported PA. If this is the case, it further strengthens the assertion of \cite{Huelamo2026} that IRAM04191 is a binary candidate.

    \item The emission from the C$^{18}$O~(2--1) line is also complicated. It shows several emission components at different PAs. One of these components traced in the velocity channels from V$_{lsr}$ = 6.9 to 7.4 km.s$^{-1}$, has morphology consistent with an even denser part of an outflow cavity (Fig. \ref{13CO_red_emission_vs_C18O_red_emission}). The PV diagram of this emission (Fig \ref{C18O_pvdiagrams}, right panels) shows kinematic signatures associated with out-flowing gas. The other emission components detected in the other velocity channels seem to have a mixed origin. Some may be associated to the outflow found by \cite{Podio2021} and some may be linked to disc emission or may be part of a rotating envelope. 

   \item Rotating gas seems to be present around IRAM04191. This is based on the double peaked profiles of $^{13}$CO~(3--2), C$^{18}$O~(2--1) and SO~6$_{5}$--5$_{4}$. The PV diagram in SO~6$_{5}$--5$_{4}$ shows a velocity gradient in the east-west direction consistent with gas rotating anti-clockwise. This rotation direction is opposite to that of the rotating envelope and hence, may strengthen our argument that this structure is potentially part of a circumstellar disc around IRAM04191, which has already been inferred using C$^{18}$O (2--1) observations \citep{Maury2020,Huelamo2026}. The PV diagram in C$^{18}$O~(2--1) at a PA of 125$^{\circ}$ also shows a signature that may hint towards rotating gas. We note, however, that a Keplerian profile cannot be determined with the current angular resolution of our data. We suggest that the rotating gas traced by the SO~6$_{5}$--5$_{4}$ could be a yet uncollapsed part of the disc detected in C$^{18}$O that is not yet in Keplerian motion. 
\end{enumerate}
\noindent
In conclusion, the results presented here clearly show that IRAM04191 has a complex formation history. There is evidence suggesting that both mass accretion and mass-loss processes are occurring simultaneously. If these mass accretion and mass-loss processes follow the relationships found for protostars, but scaled down, then it may be concluded that IRAM04191 formed in via a star-like formation pathway. \citet{Palau2024} has shown for IRAM04191 that both the infall mass (related to the mass accrertion processes) and the radio jet, detected at 3.6 cm (related to mass-loss processes) follow the relationship found for protostars (see their Fig. 5). This indicates that IRAM04191 most probably formed like a star. Despite this, we cannot conclude on the fate of this object after the Class 0 stage. 
\newline\par\noindent
Further dedicated ALMA observations on this source covering the intermediate spatial scales ($\sim$ few 1000s of au) in between the spatial scale of the results presented here and the large scale structures identified by the literature are needed to fully understand the complex web of structures that we have observed for IRAM04191 in relation to the large scale phenomena (e.g. the large 0.1 pc outflow seen in CO). These observations will in particular help with constraining both the mass loss and mass accretion processes ongoing in this source.      
\newline\par\noindent
In this context, to understand if the majority of BDs are formed as stars, more sub-stellar candidates in very early formation stages need to be studied in detail. VeLLOs such as IRAM04191 may prove to be promising sub-stellar candidates and hence warrant further study. Future James Webb Space Telescope (JWST) observations will be critical in finding and characterising these VeLLO objects to determine if they are sub-stellar. 

\bibliographystyle{aa}


\begin{thebibliography}{58}
\expandafter\ifx\csname natexlab\endcsname\relax\def\natexlab#1{#1}\fi

\bibitem[{{Alves} {et~al.}(2017){Alves}, {Girart}, {Caselli}, {Franco}, {Zhao}, {Vlemmings}, {Evans}, \& {Ricci}}]{Alves2017}
{Alves}, F.~O., {Girart}, J.~M., {Caselli}, P., {et~al.} 2017, \aap, 603, L3

\bibitem[{{Anderl} {et~al.}(2020){Anderl}, {Maret}, {Cabrit}, {Maury}, {Belloche}, {Andr{\'e}}, {Bacmann}, {Codella}, {Podio}, \& {Gueth}}]{Anderl2020}
{Anderl}, S., {Maret}, S., {Cabrit}, S., {et~al.} 2020, \aap, 643, A123

\bibitem[{{Andr{\'e}} {et~al.}(1999){Andr{\'e}}, {Motte}, \& {Bacmann}}]{Andre1999}
{Andr{\'e}}, P., {Motte}, F., \& {Bacmann}, A. 1999, \apjl, 513, L57

\bibitem[{Arce {et~al.}(2013)Arce, Mardones, Corder, Garay, Noriega-Crespo, \& Raga}]{Arce2013}
Arce, H.~G., Mardones, D., Corder, S.~A., {et~al.} 2013, The Astrophysical Journal, 774, 39

\bibitem[{{Areal} {et~al.}(2018){Areal}, {Paron}, {Celis Pe{\~n}a}, \& {Ortega}}]{Areal2018}
{Areal}, M.~B., {Paron}, S., {Celis Pe{\~n}a}, M., \& {Ortega}, M.~E. 2018, \aap, 612, A117

\bibitem[{{Bally}(2016)}]{Bally2016}
{Bally}, J. 2016, \araa, 54, 491

\bibitem[{{Belloche} {et~al.}(2002){Belloche}, {Andr{\'e}}, {Despois}, \& {Blinder}}]{Belloche2002}
{Belloche}, A., {Andr{\'e}}, P., {Despois}, D., \& {Blinder}, S. 2002, \aap, 393, 927

\bibitem[{{Birney} {et~al.}(2024){Birney}, {Dougados}, {Whelan}, {Nisini}, {Cabrit}, \& {Zhang}}]{Birney2024}
{Birney}, M., {Dougados}, C., {Whelan}, E.~T., {et~al.} 2024, \aap, 692, A143

\bibitem[{Chabrier \& Lenoble(2023)}]{Chabrier2023}
Chabrier, G. \& Lenoble, R. 2023, The Astrophysical Journal Letters, 944, L33

\bibitem[{Comrie {et~al.}(2021)Comrie, Wang, Hsu, Moraghan, Harris, Pang, Pińska, Chiang, Chang, Hwang, Jan, Lin, \& Simmonds}]{Comrie2021}
Comrie, A., Wang, K.-S., Hsu, S.-C., {et~al.} 2021, CARTA: The Cube Analysis and Rendering Tool for Astronomy

\bibitem[{{di Francesco} {et~al.}(2007){di Francesco}, {Evans}, {Caselli}, {Myers}, {Shirley}, {Aikawa}, \& {Tafalla}}]{diFrancesco2007}
{di Francesco}, J., {Evans}, II, N.~J., {Caselli}, P., {et~al.} 2007, in Protostars and Planets V, ed. B.~{Reipurth}, D.~{Jewitt}, \& K.~{Keil}, 17

\bibitem[{{Dunham} {et~al.}(2006){Dunham}, {Evans}, {Bourke}, {Dullemond}, {Young}, {Brooke}, {Chapman}, {Myers}, {Porras}, {Spiesman}, {Teuben}, \& {Wahhaj}}]{Dunham2006}
{Dunham}, M.~M., {Evans}, II, N.~J., {Bourke}, T.~L., {et~al.} 2006, \apj, 651, 945

\bibitem[{{Dutrey} {et~al.}(2024){Dutrey}, {Chapillon}, {Guilloteau}, {Tang}, {Boccaletti}, {Bouscasse}, {Collin-Dufresne}, {Di Folco}, {Fuente}, {Pi{\'e}tu}, {Rivi{\`e}re-Marichalar}, \& {Semenov}}]{Dutrey2024}
{Dutrey}, A., {Chapillon}, E., {Guilloteau}, S., {et~al.} 2024, \aap, 689, L7

\bibitem[{{Eisl{\"o}ffel}(2000)}]{Eisloffel2000}
{Eisl{\"o}ffel}, J. 2000, \aap, 354, 236

\bibitem[{{Evans} {et~al.}(2009){Evans}, {Dunham}, {J{\o}rgensen}, {Enoch}, {Mer{\'\i}n}, {van Dishoeck}, {Alcal{\'a}}, {Myers}, {Stapelfeldt}, {Huard}, {Allen}, {Harvey}, {van Kempen}, {Blake}, {Koerner}, {Mundy}, {Padgett}, \& {Sargent}}]{Evans2009}
{Evans}, II, N.~J., {Dunham}, M.~M., {J{\o}rgensen}, J.~K., {et~al.} 2009, \apjs, 181, 321

\bibitem[{{Ferreira}(1997)}]{Ferreira1997}
{Ferreira}, J. 1997, \aap, 319, 340

\bibitem[{{Fischer} {et~al.}(2023){Fischer}, {Hillenbrand}, {Herczeg}, {Johnstone}, {Kospal}, \& {Dunham}}]{Fischer2023}
{Fischer}, W.~J., {Hillenbrand}, L.~A., {Herczeg}, G.~J., {et~al.} 2023, in Astronomical Society of the Pacific Conference Series, Vol. 534, Protostars and Planets VII, ed. S.~{Inutsuka}, Y.~{Aikawa}, T.~{Muto}, K.~{Tomida}, \& M.~{Tamura}, 355

\bibitem[{{Flores} {et~al.}(2023){Flores}, {Ohashi}, {Tobin}, {J{\o}rgensen}, {Takakuwa}, {Li}, {Lin}, {van't Hoff}, {Plunkett}, {Yamato}, {Sai (Insa Choi)}, {Koch}, {Yen}, {Aikawa}, {Aso}, {de Gregorio-Monsalvo}, {Kido}, {Kwon}, {Lee}, {Lee}, {Looney}, {Santamar{\'\i}a-Miranda}, {Sharma}, {Thieme}, {Williams}, {Han}, {Narayanan}, \& {Lai}}]{Flores2023}
{Flores}, C., {Ohashi}, N., {Tobin}, J.~J., {et~al.} 2023, \apj, 958, 98

\bibitem[{{Frank} {et~al.}(2014){Frank}, {Ray}, {Cabrit}, {Hartigan}, {Arce}, {Bacciotti}, {Bally}, {Benisty}, {Eisl{\"o}ffel}, {G{\"u}del}, {Lebedev}, {Nisini}, \& {Raga}}]{Frank2014}
{Frank}, A., {Ray}, T.~P., {Cabrit}, S., {et~al.} 2014, in Protostars and Planets VI, ed. H.~{Beuther}, R.~S. {Klessen}, C.~P. {Dullemond}, \& T.~{Henning}, 451--474

\bibitem[{{Garufi} {et~al.}(2022){Garufi}, {Podio}, {Codella}, {Segura-Cox}, {Vander Donckt}, {Mercimek}, {Bacciotti}, {Fedele}, {Kasper}, {Pineda}, {Humphreys}, \& {Testi}}]{Garufi2022}
{Garufi}, A., {Podio}, L., {Codella}, C., {et~al.} 2022, \aap, 658, A104

\bibitem[{{Hales} {et~al.}(2024){Hales}, {Gupta}, {Ru{\'\i}z-Rodr{\'\i}guez}, {Williams}, {P{\'e}rez}, {Cieza}, {Gonz{\'a}lez-Ruilova}, {Pineda}, {Santamar{\'\i}a-Miranda}, {Tobin}, {Weber}, {Zhu}, \& {Zurlo}}]{Hales2024}
{Hales}, A.~S., {Gupta}, A., {Ru{\'\i}z-Rodr{\'\i}guez}, D., {et~al.} 2024, \apj, 966, 96

\bibitem[{{Huélamo} {et~al.}(2026){Huélamo}, {de Gregorio-Monsalvo, I.}, {Palau, Aina}, {Carrasco-González, C.}, {Ribas, A.}, {Bouy, H.}, {Pandey, R.}, {Barrado, D.}, {Otten, N.}, {Ivanov, V. D.}, {Sterzik, M. F.}, {Dunham, M.}, {Zapata, L. A.}, {Pantin, E.}, \& {Macias, E.}}]{Huelamo2026}
{Huélamo}, {de Gregorio-Monsalvo, I.}, {Palau, Aina}, {et~al.} 2026, \aapƒ, 709, L3

\bibitem[{{J{\o}rgensen} {et~al.}(2013){J{\o}rgensen}, {Visser}, {Sakai}, {Bergin}, {Brinch}, {Harsono}, {Lindberg}, {van Dishoeck}, {Yamamoto}, {Bisschop}, \& {Persson}}]{Jorgensen2013}
{J{\o}rgensen}, J.~K., {Visser}, R., {Sakai}, N., {et~al.} 2013, \apjl, 779, L22

\bibitem[{{Kido} {et~al.}(2023){Kido}, {Takakuwa}, {Saigo}, {Ohashi}, {Tobin}, {J{\o}rgensen}, {Aikawa}, {Aso}, {Encalada}, {Flores}, {Gavino}, {de Gregorio-Monsalvo}, {Han}, {Hirano}, {Koch}, {Kwon}, {Lai}, {Lee}, {Lee}, {Li}, {Lin}, {Looney}, {Mori}, {Narayanan}, {Plunkett}, {Phuong}, {(Insa Choi)}, {Santamar{\'\i}a-Miranda}, {Sharma}, {Sheehan}, {Thieme}, {Tomida}, {van't Hoff}, {Williams}, {Yamato}, \& {Yen}}]{Kido2023}
{Kido}, M., {Takakuwa}, S., {Saigo}, K., {et~al.} 2023, \apj, 953, 190

\bibitem[{{Kim} {et~al.}(2019){Kim}, {Lee}, {Maheswar}, {Kim}, {Soam}, {Saito}, {Kiyokane}, \& {Kim}}]{Kim2019}
{Kim}, G., {Lee}, C.~W., {Maheswar}, G., {et~al.} 2019, \apjs, 240, 18

\bibitem[{Lee {et~al.}(2010)Lee, Hasegawa, Hirano, Palau, Shang, Ho, \& Zhang}]{CFLee2010}
Lee, C.-F., Hasegawa, T.~I., Hirano, N., {et~al.} 2010, The Astrophysical Journal, 713, 731

\bibitem[{{Lee} {et~al.}(2005){Lee}, {Ho}, \& {White}}]{CFLee20052005}
{Lee}, C.-F., {Ho}, P. T.~P., \& {White}, S.~M. 2005, \apj, 619, 948

\bibitem[{{Lee} {et~al.}(2002){Lee}, {Mundy}, {Stone}, \& {Ostriker}}]{CFLee2002}
{Lee}, C.-F., {Mundy}, L.~G., {Stone}, J.~M., \& {Ostriker}, E.~C. 2002, \apj, 576, 294

\bibitem[{{Maury}(2020)}]{Maury2020}
{Maury}, A. 2020, in IAU Symposium, Vol. 345, Origins: From the Protosun to the First Steps of Life, ed. B.~G. {Elmegreen}, L.~V. {T{\'o}th}, \& M.~{G{\"u}del}, 91--95

\bibitem[{{Maury} {et~al.}(2019){Maury}, {Andr{\'e}}, {Testi}, {Maret}, {Belloche}, {Hennebelle}, {Cabrit}, {Codella}, {Gueth}, {Podio}, {Anderl}, {Bacmann}, {Bontemps}, {Gaudel}, {Ladjelate}, {Lef{\`e}vre}, {Tabone}, \& {Lefloch}}]{Maury2019}
{Maury}, A.~J., {Andr{\'e}}, P., {Testi}, L., {et~al.} 2019, \aap, 621, A76

\bibitem[{{McMullin} {et~al.}(2007){McMullin}, {Waters}, {Schiebel}, {Young}, \& {Golap}}]{McMullin2007}
{McMullin}, J.~P., {Waters}, B., {Schiebel}, D., {Young}, W., \& {Golap}, K. 2007, in Astronomical Society of the Pacific Conference Series, Vol. 376, Astronomical Data Analysis Software and Systems XVI, ed. R.~A. {Shaw}, F.~{Hill}, \& D.~J. {Bell}, 127

\bibitem[{{Neupane} {et~al.}(2024){Neupane}, {Wyrowski}, {Menten}, {Urquhart}, {Colombo}, {Lin}, \& {Garay}}]{Neupane2024}
{Neupane}, S., {Wyrowski}, F., {Menten}, K.~M., {et~al.} 2024, \aap, 692, A114

\bibitem[{Noriega-Crespo {et~al.}(2014)Noriega-Crespo, Raga, Moro-Martín, Flagey, \& Carey}]{Noriega-Crespo_2014}
Noriega-Crespo, A., Raga, A.~C., Moro-Martín, A., Flagey, N., \& Carey, S.~J. 2014, New Journal of Physics, 16, 105008

\bibitem[{{Padoan} \& {Nordlund}(2002)}]{PadoanANDNordlund2002}
{Padoan}, P. \& {Nordlund}, {\r{A}}. 2002, \apj, 576, 870

\bibitem[{{Padoan} \& {Nordlund}(2004)}]{PadoanANDNordlund2004}
{Padoan}, P. \& {Nordlund}, {\r{A}}. 2004, \apj, 617, 559

\bibitem[{{Palau} {et~al.}(2024){Palau}, {Hu{\'e}lamo}, {Barrado}, {Dunham}, \& {Lee}}]{Palau2024}
{Palau}, A., {Hu{\'e}lamo}, N., {Barrado}, D., {Dunham}, M.~M., \& {Lee}, C.~W. 2024, \nar, 99, 101711

\bibitem[{{Pascucci} {et~al.}(2025){Pascucci}, {Beck}, {Cabrit}, {Bajaj}, {Edwards}, {Louvet}, {Najita}, {Skinner}, {Gorti}, {Salyk}, {Brittain}, {Krijt}, {Muzerolle Page}, {Ruaud}, {Schwarz}, {Semenov}, {Duch{\^e}ne}, \& {Villenave}}]{Pascucci2025}
{Pascucci}, I., {Beck}, T.~L., {Cabrit}, S., {et~al.} 2025, Nature Astronomy, 9, 81

\bibitem[{{Pascucci} {et~al.}(2023){Pascucci}, {Cabrit}, {Edwards}, {Gorti}, {Gressel}, \& {Suzuki}}]{Pascucci2023}
{Pascucci}, I., {Cabrit}, S., {Edwards}, S., {et~al.} 2023, in Astronomical Society of the Pacific Conference Series, Vol. 534, Protostars and Planets VII, ed. S.~{Inutsuka}, Y.~{Aikawa}, T.~{Muto}, K.~{Tomida}, \& M.~{Tamura}, 567

\bibitem[{{Pineda} {et~al.}(2020){Pineda}, {Segura-Cox}, {Caselli}, {Cunningham}, {Zhao}, {Schmiedeke}, {Maureira}, \& {Neri}}]{Pineda2020}
{Pineda}, J.~E., {Segura-Cox}, D., {Caselli}, P., {et~al.} 2020, Nature Astronomy, 4, 1158

\bibitem[{{Podio} {et~al.}(2021){Podio}, {Tabone}, {Codella}, {Gueth}, {Maury}, {Cabrit}, {Lefloch}, {Maret}, {Belloche}, {Andr{\'e}}, {Anderl}, {Gaudel}, \& {Testi}}]{Podio2021}
{Podio}, L., {Tabone}, B., {Codella}, C., {et~al.} 2021, \aap, 648, A45

\bibitem[{Ray \& Ferreira(2021)}]{Ray2021}
Ray, T. \& Ferreira, J. 2021, New Astronomy Reviews, 93, 101615

\bibitem[{Reipurth \& Mikkola(2015)}]{Reipurth2015}
Reipurth, B. \& Mikkola, S. 2015, The Astronomical Journal, 149, 145

\bibitem[{{Riaz} {et~al.}(2017){Riaz}, {Brice{\~n}o}, {Whelan}, \& {Heathcote}}]{Riaz2017}
{Riaz}, B., {Brice{\~n}o}, C., {Whelan}, E.~T., \& {Heathcote}, S. 2017, \apj, 844, 47

\bibitem[{Riaz {et~al.}(2024)Riaz, Stamatellos, \& MacHida}]{Riaz2024ObservationsDwarf}
Riaz, B., Stamatellos, D., \& MacHida, M.~N. 2024, Monthly Notices of the Royal Astronomical Society, 529, 3601

\bibitem[{Ruaud {et~al.}(2022)Ruaud, Gorti, \& Hollenbach}]{Ruaud2022}
Ruaud, M., Gorti, U., \& Hollenbach, D.~J. 2022, The Astrophysical Journal, 925, 49

\bibitem[{{Santamar{\'\i}a-Miranda} {et~al.}(2020){Santamar{\'\i}a-Miranda}, {de Gregorio-Monsalvo}, {Hu{\'e}lamo}, {Plunkett}, {Ribas}, {Comer{\'o}n}, {Schreiber}, {L{\'o}pez}, {Mu{\v{z}}i{\'c}}, \& {Testi}}]{Santamaria2020}
{Santamar{\'\i}a-Miranda}, A., {de Gregorio-Monsalvo}, I., {Hu{\'e}lamo}, N., {et~al.} 2020, \aap, 640, A13

\bibitem[{{Santamar{\'\i}a-Miranda} {et~al.}(2021){Santamar{\'\i}a-Miranda}, {de Gregorio-Monsalvo}, {Plunkett}, {Hu{\'e}lamo}, {L{\'o}pez}, {Ribas}, {Schreiber}, {Mu{\v{z}}i{\'c}}, {Palau}, {Knee}, {Bayo}, {Comer{\'o}n}, \& {Hales}}]{Santamaria2021}
{Santamar{\'\i}a-Miranda}, A., {de Gregorio-Monsalvo}, I., {Plunkett}, A.~L., {et~al.} 2021, \aap, 646, A10

\bibitem[{{Stahler}(1994)}]{Stahler1994}
{Stahler}, S.~W. 1994, \apj, 422, 616

\bibitem[{{Stock} {et~al.}(2020){Stock}, {Caratti o Garatti}, {McGinnis}, {Garcia Lopez}, {Antoniucci}, {Fedriani}, \& {Ray}}]{Stock2020}
{Stock}, C., {Caratti o Garatti}, A., {McGinnis}, P., {et~al.} 2020, \aap, 643, A181

\bibitem[{Takakuwa {et~al.}(2003)Takakuwa, Ohashi, \& Hirano}]{Takakuwa_2003}
Takakuwa, S., Ohashi, N., \& Hirano, N. 2003, The Astrophysical Journal, 590, 932

\bibitem[{Teague(2019)}]{GoFish}
Teague, R. 2019, The Journal of Open Source Software, 4, 1632

\bibitem[{{Tychoniec} {et~al.}(2021){Tychoniec}, {van Dishoeck}, {van't Hoff}, {van Gelder}, {Tabone}, {Chen}, {Harsono}, {Hull}, {Hogerheijde}, {Murillo}, \& {Tobin}}]{Tychoniec2021}
{Tychoniec}, {\L}., {van Dishoeck}, E.~F., {van't Hoff}, M. L.~R., {et~al.} 2021, \aap, 655, A65

\bibitem[{{Valdivia-Mena} {et~al.}(2024){Valdivia-Mena}, {Pineda}, {Caselli}, {Segura-Cox}, {Schmiedeke}, {Spezzano}, {Offner}, {Ivlev}, {Kuffmeier}, {Cunningham}, {Neri}, \& {Maureira}}]{Valdivia-Mena2024}
{Valdivia-Mena}, M.~T., {Pineda}, J.~E., {Caselli}, P., {et~al.} 2024, \aap, 687, A71

\bibitem[{{Vorobyov} {et~al.}(2017){Vorobyov}, {Elbakyan}, {Dunham}, \& {Guedel}}]{Vorobyov2017}
{Vorobyov}, E.~I., {Elbakyan}, V., {Dunham}, M.~M., \& {Guedel}, M. 2017, \aap, 600, A36

\bibitem[{{Whelan} {et~al.}(2014){Whelan}, {Alcal{\'a}}, {Bacciotti}, {Nisini}, {Bonito}, {Antoniucci}, {Stelzer}, {Biazzo}, {D'Elia}, \& {Ray}}]{Whelan2014}
{Whelan}, E.~T., {Alcal{\'a}}, J.~M., {Bacciotti}, F., {et~al.} 2014, \aap, 570, A59

\bibitem[{{Whelan} {et~al.}(2005){Whelan}, {Ray}, {Bacciotti}, {Natta}, {Testi}, \& {Randich}}]{Whelan2005}
{Whelan}, E.~T., {Ray}, T.~P., {Bacciotti}, F., {et~al.} 2005, \nat, 435, 652

\bibitem[{{Wu} {et~al.}(2004){Wu}, {Wei}, {Zhao}, {Shi}, {Yu}, {Qin}, \& {Huang}}]{Wu2004}
{Wu}, Y., {Wei}, Y., {Zhao}, M., {et~al.} 2004, \aap, 426, 503

\bibitem[{{Zhou} {et~al.}(2023){Zhou}, {Wyrowski}, {Neupane}, {Urquhart}, {Evans}, {V{\'a}zquez-Semadeni}, {Menten}, {Gong}, \& {Liu}}]{Zhou2023}
{Zhou}, J.~W., {Wyrowski}, F., {Neupane}, S., {et~al.} 2023, \aap, 676, A69

\end{thebibliography}

\begin{acknowledgements}
N. Otten acknowledges previous funding from the Maynooth University Graduate Teaching Scholarship \& Taighde Éireann (Research Ireland) under the RI-ESO Studentship Agreement for this work and is currently funded by Research Ireland though grant 21/PATH-S/9360(T) (PI: P. Kavanagh).

N. Huélamo is funded by the Spanish grant MCIN/AEI/10.13039/501100011033 PID2023-150468NB-I00.

G.Blazquez.-Calero acknowledges financial support from grants PID2023-146295NB-I00 and CEX2021-001131-S, funded by MCIN/AEI/10.13039/501100011033, from Junta de Andalucía (Spain) grant P20-00880 (FEDER, EU), and ESO Science Support Discretionary Fund for their financial support under the 2024 SSDF 06 project.
\end{acknowledgements}

\onecolumn
\begin{appendix}

\section{Summary of Observations}

This appendix contains two tables which summarise the archival observations and the properties of the final emission line image cubes used to produce the figures presented in the main body. 

\begin{table}[!htbp]
    \centering
    \caption{Summary of ALMA observations.}
    \resizebox{\textwidth}{!}{%
    \begin{tabular}{ccccccccccccc}\hline\hline
         Array & Date & Project ID & PWV & Flux Cal. & Phase Cal. & Bandpass Cal. & Avg. Elevation & Intr. Time & AR & MRS & Baseline (min-max)  \\
         & [dd-mm-yyyy] & & [mm] & & & &[$^{\circ}$] & [s] & [arcsec] & [arcsec] & [m] \\
         \hline
         12m & 15-10-2016 & 2016.1.01284.S & 0.641 & J0510+1800 & J0433+0521 & J0425+1755 & 37.5 & 3991.680 & 0.154 & 1.949 & 18 - 1809 \\
         12m & 07-10-2019 & 2019.1.00847.S & 0.436 & J0510+1800 & J0431+1731 & J0431+2037 & 46.7 & 907.200 & 0.285 & 3.758 & 15 - 1547 \\
         \hline
    \end{tabular}
    }
    \tablefoot{AR = Angular Resolution. AR is defined as the smallest scale structure that can be resolved by the interferometer. This is determined by the length of the longest baseline and the wavelength of the observation. MRS = Maximum Recoverable Scale. MRS is defined as the largest scale structure that can be traced by the interferometer. This is determined by the length of the shortest baseline and the wavelength of the observation.} 
    \label{alma_obslog}
\end{table}

\vspace{0.5cm}

\begin{table}[!htbp]
    \centering\small
    \caption{Properties of the produced emission line cubes.}
    \resizebox{\textwidth}{!}{%
    \begin{tabular}{cccccccccccc} \hline\hline
        Species  & Transition & Robust & RA range & Dec range & Beam Major Axis & Beam Minor Axis & Beam PA & RMS & Spectral Res. \\
         & & & [J2000] & [J2000] & [arcsec] & [arcsec] & [$^{\circ}$] & [mJy/beam] & [km.s$^{-1}$]\\
         \hline
         $^{13}$CO & 3--2 & 2.0 & 04:21:56.076, 04:21:57.734 & +15.29.34.049, +15.29.58.009 & 0.39 & 0.33 & 155.6 & 9.48 & 0.221 \\
         C$^{18}$O & 2--1 & 0.5  & 04:21:56.739, 04:21:57.022 & +15.29.43.952, +15.29.48.044 & 0.19 & 0.17 & 102.1 & 2.62 & 0.083 \\
         SO & 6$_{5}$--5$_{4}$ & 1.0 & 04:21:56.548, 04:21:57.255 & +15.29.40.990, +15.29.51.210 & 0.24 & 0.20 & 119.4 & 2.10 & 0.083 \\
         \hline
    \end{tabular}
    }
    \tablefoot{RA range  = Right Ascension extent, Dec range = Declination extent, PA = Position Angle (measured east of north), RMS = Root mean square noise,  Res. = Resolution.}
    \label{beam_params_table}
\end{table}

\twocolumn
\section{Description of the molecular lines used for this study.}
In this appendix, we present our reasoning for choosing to focus on the $^{13}$CO~(3--2), C$^{18}$O~(2--1) and  SO~6$_{5}$--5$_{4}$ molecular emission lines for this study. These lines were chosen based on their ability to trace cold/dense gas in close proximity of IRAM04191 (from a few 10s to several 100s of au). The transition selected for each line was chosen based on the available public archival 12m array data for IRAM04191. We now provide a brief description of what each molecular line presented in this article can trace according to current literature. \\

\noindent The $^{13}$CO~(3--2) emission line traces dense ($\sim$ $\times$ 10$^{3}$ - 10$^{6}$ cm$^{-3}$) and warm (few 10s to 100 K) gas. $^{13}$CO~(3--2) is an isotopologue of the more abundant CO molecule and as such, it traces denser gas closer to the emitting source than CO. The literature has shown that transitions of $^{13}$CO emission have been found in molecular outflows emanating from very low mass stars \citep{Santamaria2020} and in thin filamentary structures within molecular clouds \citep{Neupane2024}. \\

\noindent C$^{18}$O~(2--1) is another isotopologue of CO and a tracer of gas at temperatures of $\sim$ 15 K. C$^{18}$O~(2--1) is less abundant than CO and $^{13}$CO hence, it traces even denser gas closer to the emitting source. The 2--1 transition of C$^{18}$O has an upper level energy of  $\sim$ 15 K (Table \ref{line_table}) while the 3--2 transition of C$^{18}$O has an upper level energy of  $\sim$ 32 K. This makes our selected transition of C$^{18}$O particularly useful for detecting cold gas in close proximity to our target. The literature has shown that transitions of C$^{18}$O gas emission have been observed in several different structures around low mass YSOs and BDs. It has been found in the cold envelope regions surrounding young stellar systems \citep{Tychoniec2021} and in the discs surrounding young protostars and BDs \citep{Alves2017,Ruaud2022}. \\

\noindent The SO~6$_{5}$--5$_{4}$ molecular emission line has been detected in shocked regions, close to the circumstellar disc \citep{Dutrey2024}. Emission from oxygen bearing species such as SO have also been observed in the knots within the high velocity molecular jet launched by low mass young stars and in the jet bow shocks \citep{Tychoniec2021,CFLee2010,Frank2014}. This molecule is also associated to streamers, as either infalling rotating structures \citep{Flores2023} or in shocks where the streamer impacts flattened disc like structures \citep{Kido2023}. 

\begin{table}[tbh!]
    \centering
    \caption{Properties of the emission lines used in this study. Taken from the Advanced Splatalogue database (\url{https://splatalogue.online/\#/advanced}). }
    \begin{tabular}{lcccc} \hline\hline
        Species & Transition & Rest Frequency & \textit{E$_\mathrm{low}$} & \textit{E$_\mathrm{up}$} \\
         & & [GHz] & [K] & [K]\\
         \hline
         $^{13}$CO  & 3--2 & 330.588 & 15.9 & 31.7 \\
         C$^{18}$O & 2--1 & 219.560 & 5.3 & 15.8 \\
         SO & 6$_{5}$--5$_{4}$ & 219.949 & 24.4 & 35.0 \\
         
         \hline
    \end{tabular}
    \label{line_table}
      
\end{table} 
\section{Channel Maps}\label{appendix_channel_maps}
This appendix contains the channel maps generated from the $^{13}$CO~(3--2), C$^{18}$O~(2--1) and SO~6$_{5}$--5$_{4}$ image cubes using the \texttt{Go Fish} \citep{GoFish} package in \textbf{\textit{Python}}. \\
\clearpage
\noindent\begin{minipage}{\textwidth}
    \centering
    
    \includegraphics[height=0.89\vsize, width=\hsize, keepaspectratio = True]{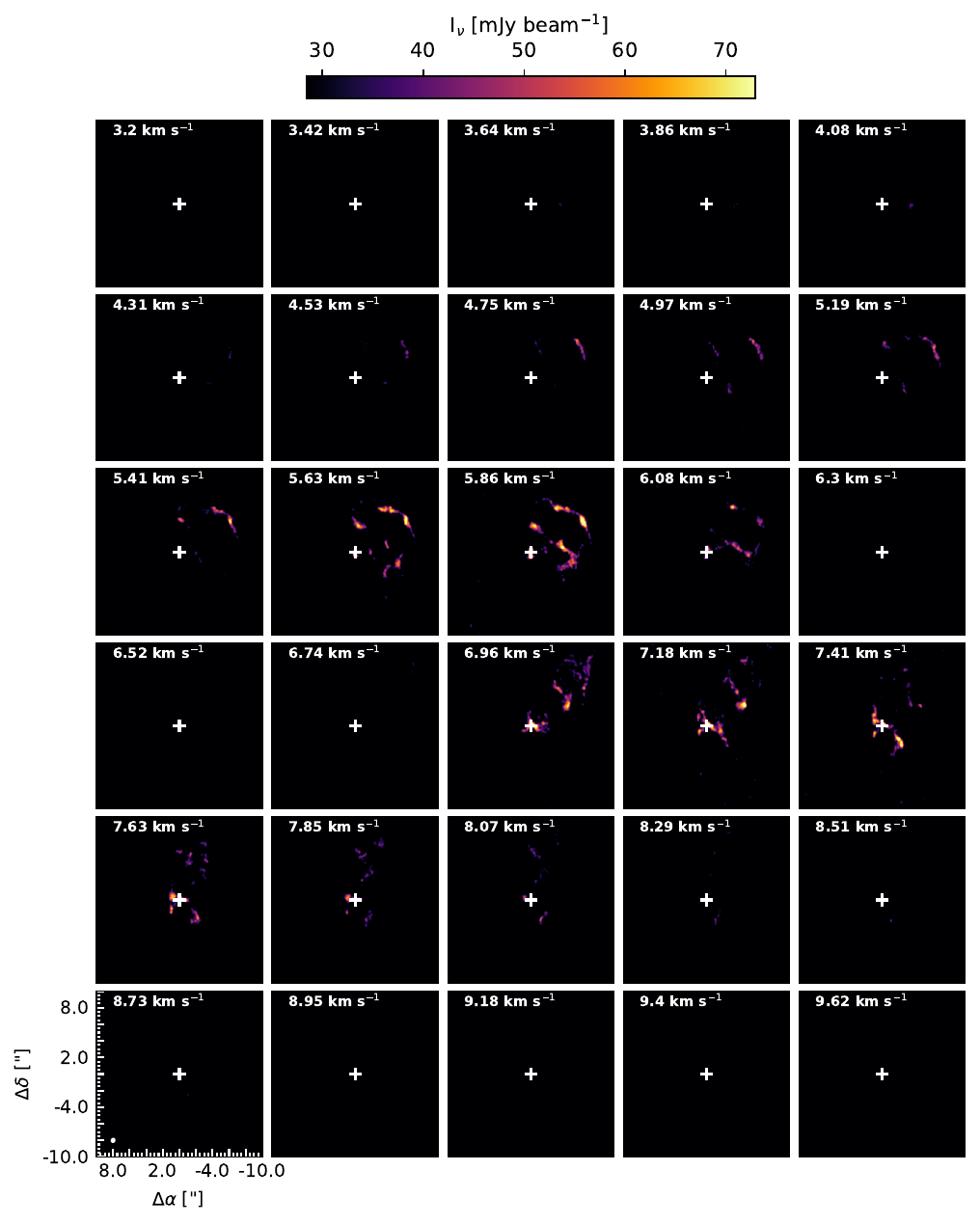}
    \captionof{figure}{Channel maps for $^{13}$CO~(3--2) emission in IRAM04191+1522 from V$_{lsr}$ = 3.2 km.s$^{-1}$ to 9.62 km.s$^{-1}$. The emission shown is clipped at 3$\sigma_\mathrm{rms}$ and above ($\sigma_\mathrm{rms}$ = 9.48 mJy/beam). The beam (0$\farcs$39 $\times$ 0$\farcs$33 at a PA of 155.6 $^{\circ}$) is shown in the bottom left of the lower left channel map. The systemic velocity is 6.5~km.s$^{-1}$. The source position is indicated by the white cross.}
    \label{13CO_channel_maps}
\end{minipage}

\begin{figure*}
    \centering
    \includegraphics[height = \vsize, width = \hsize, keepaspectratio = True]{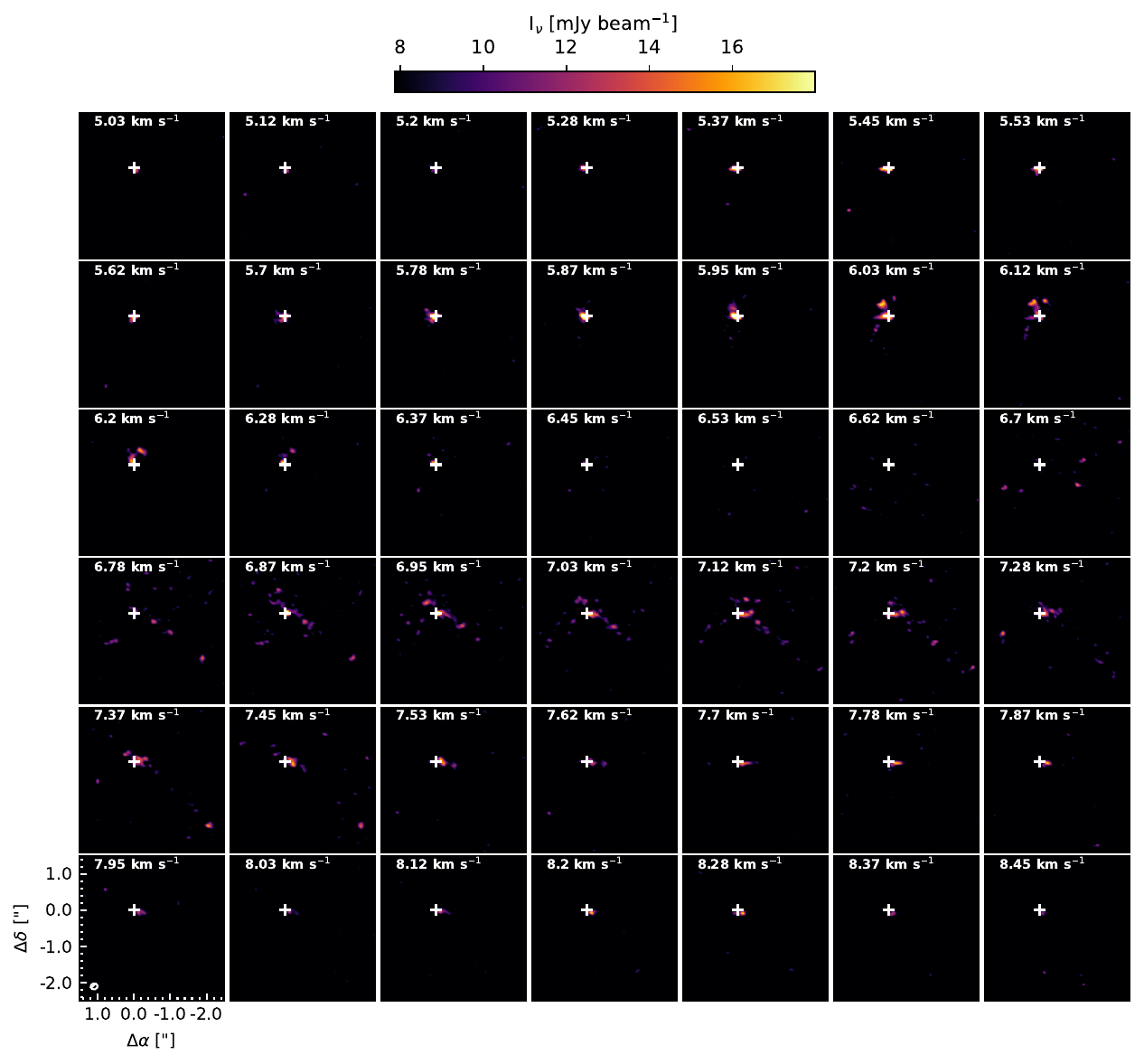}
    \caption{Channel maps for C$^{18}$O~(2--1) emission in IRAM04191+1522 from V$_{lsr}$ = 5.03 km.s$^{-1}$ to 8.45 km.s$^{-1}$. The emission shown is clipped at 3$\sigma_\mathrm{rms}$ and above ($\sigma_\mathrm{rms}$ = 2.62 mJy/beam). The beam (0$\farcs$19 $\times$ 0$\farcs$17 at a PA of 102.1 $^{\circ}$) is shown in the bottom left of the lower left channel map. The systemic velocity is 6.5~km.s$^{-1}$. The source position is indicated by the white cross.}
    \label{C18O_channel_maps}
\end{figure*}

\begin{figure*}
    \centering
    \includegraphics[width = 0.92\linewidth, keepaspectratio = True]{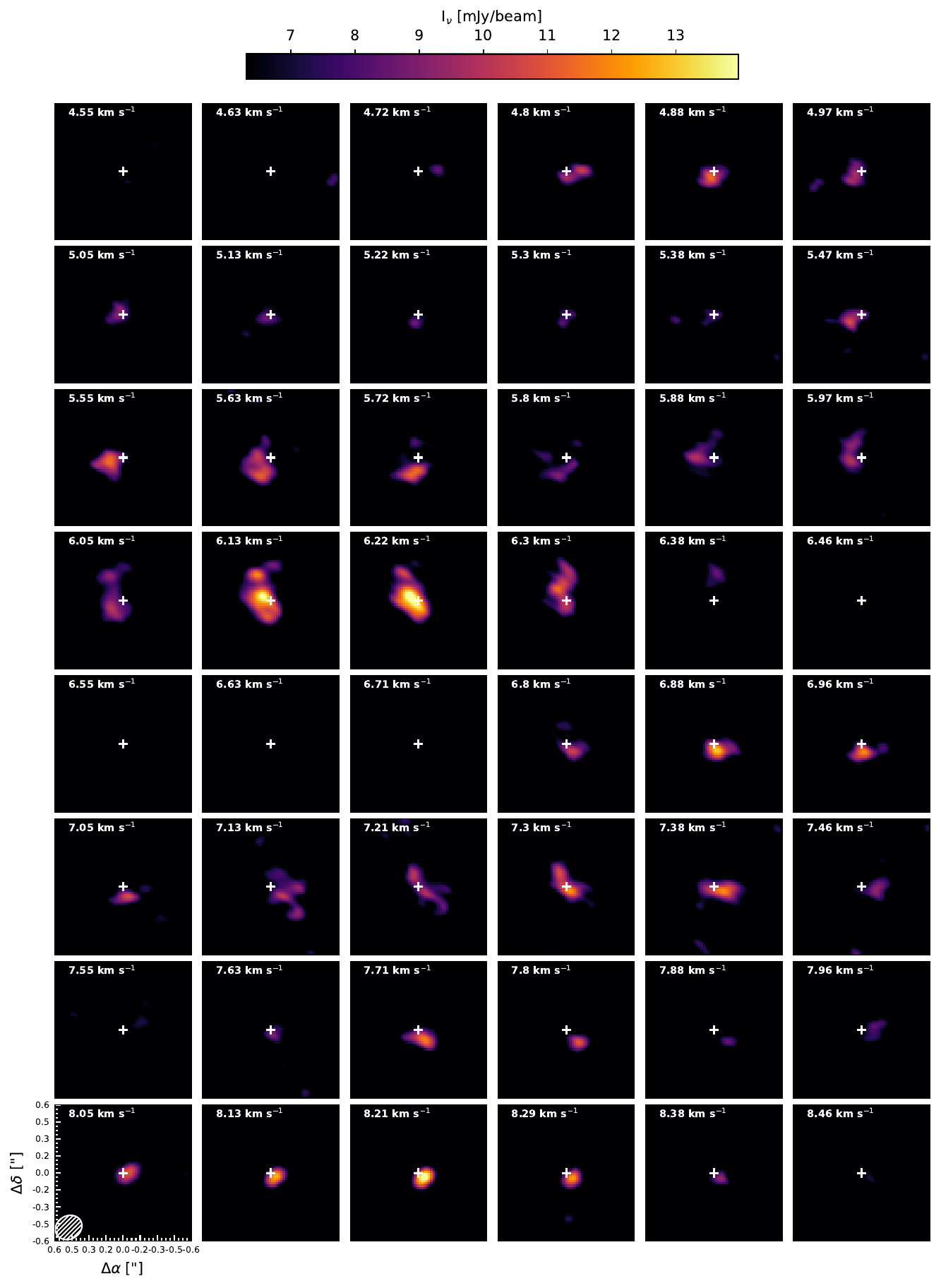}
    \caption{Channel maps for SO~6$_{5}$--5$_{4}$ emission in IRAM04191+1522 from V$_{lsr}$ = 4.55 km.s$^{-1}$ to 8.46 km.s$^{-1}$. The emission shown is clipped at 3$\sigma_\mathrm{rms}$ and above ($\sigma_\mathrm{rms}$ = 2.10 mJy/beam) The beam (0$\farcs$24 $\times$ 0$\farcs$20 at a PA of 119.4 $^{\circ}$) is shown in the bottom left of the lower left channel map. The systemic velocity is 6.5~km.s$^{-1}$. The source position is indicated by the white cross.}
    \label{SO_channel_maps}
\end{figure*}

\onecolumn
\section{The outflow cavity: PV diagram for $^{13}$CO compared to C$^{18}$O and extracted cuts.}\label{AppendixD}
In this appendix we present the PV diagram of $^{13}$CO (3--2) taken along the same PA as the purple pseudo slit shown in the left panel of Fig. \ref{C18O_pvdiagrams}, and compare it with the PV diagram presented in the top right panel of Fig. \ref{C18O_pvdiagrams}. Additionally, we take cuts along each of these PV diagrams extracted at a PA of 54$^{\circ}$ in both of these emission lines to show the presence of outflowing gas at a PA of 234$^{\circ}$. 

\begin{figure}[tbh!]
    \centering
    \includegraphics[width=0.7\linewidth]{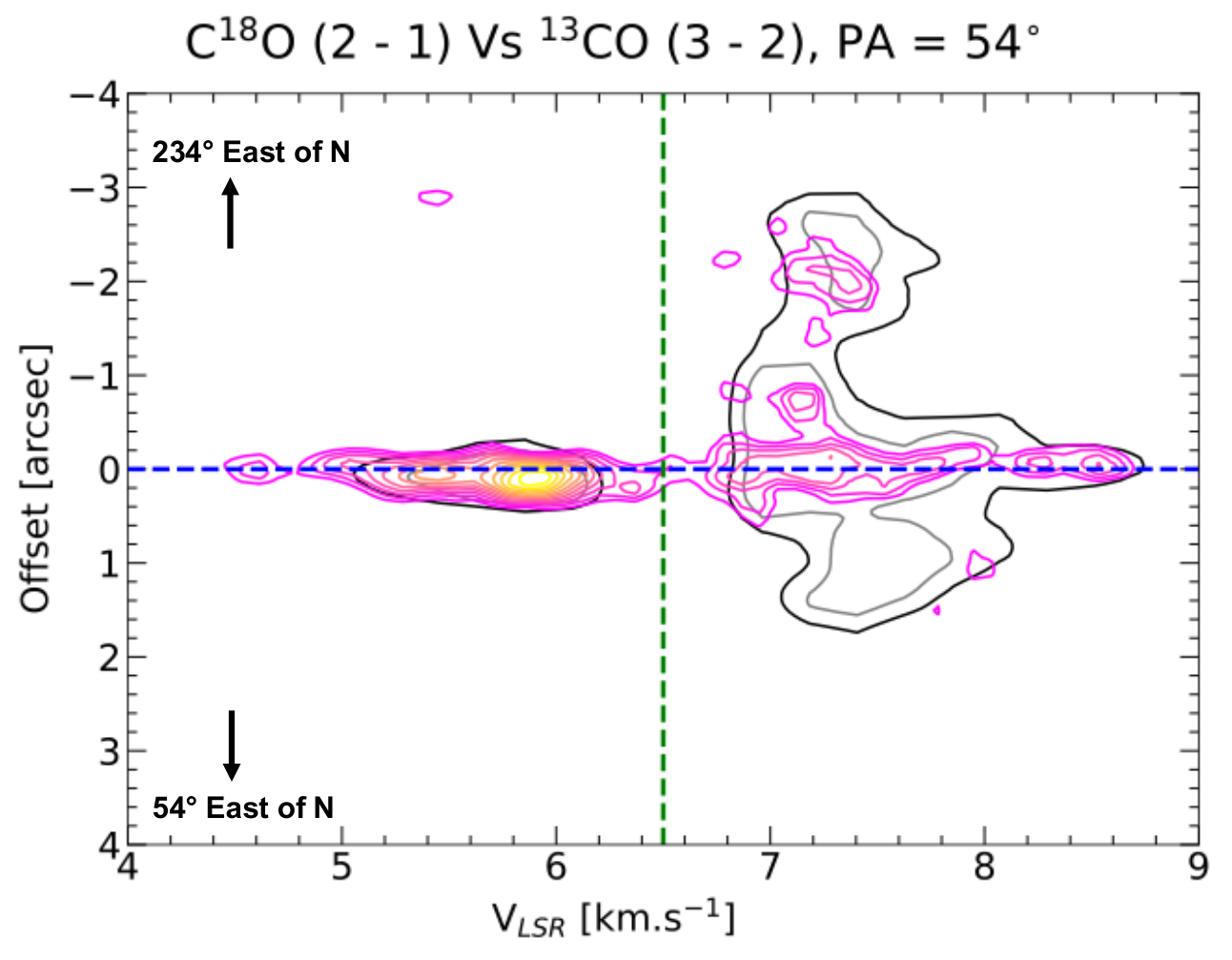}
    \caption{Comparison of the C$^{18}$O PV diagram taken at a PA of 54$^{\circ}$ (as shown in the top right panel of Fig. \ref{C18O_pvdiagrams}) in pink/yellow contours to a corresponding PV diagram extracted at the same PA for the $^{13}$CO emission shown in grey contours. Both PV diagrams have the used the same pseudo-slit width of 0$\farcs$4 (20 pixels for C$^{18}$O and 10 pixels for $^{13}$CO). The $^{13}$CO pseudo-slit was 6$\arcsec$ long (to ensure that all the emission along the PA of 54$^{\circ}$ was captured) compared to a 3$\arcsec$ long pseudo-slit for C$^{18}$O. The contour levels used for the C$^{18}$O diagram are the same ones as for the top panel of Fig. \ref{C18O_pvdiagrams}. The contour levels used for the  $^{13}$CO PV diagram were 3 and 4$\sigma$ ($\sigma$ = 14.97 mJy beam$^{-1}$). The green and blue dashed lines correspond to the systemic velocity (see Table \ref{properties table}) and zero offset from the source position respectively. }
    \label{PVD_13co_vs_c18o_pa54}
\end{figure}

\noindent From Fig. \ref{PVD_13co_vs_c18o_pa54} it is clear that the PV diagrams for the C$^{18}$O and $^{13}$CO emission lines share a comparable morphology. Both show a feature at $\sim$ 7.5 km.s$^{-1}$ that is offset from the source by between 2 - 3$\arcsec$. The $^{13}$CO emission is slightly more extended than C$^{18}$O at this PA as $^{13}$CO is a lower density gas tracer (see Table \ref{line_table} for the properties of each line). Secondly, the peak contours of the C$^{18}$O PV diagram (at $\sim$ 6 km.s$^{-1}$) appear to align with the peak contour of the $^{13}$CO PV diagram at the same velocity. By taking cuts along these PV diagrams we show the presence of outflowing gas along a PA of 234$^{\circ}$. The cut positions and corresponding cut profiles for the $^{13}$CO and C$^{18}$O emission lines are presented in Figs. \ref{PVD_13CO_cuts} and \ref{PVD_C18O_cuts} respectively. It is noted here that the spatial resolution of the C$^{18}$O and $^{13}$CO observations are 0$\farcs$2 pixel$^{-1}$ and 0$\farcs$4 pixel$^{-1}$ respectively. As such the PV diagram of $^{13}$CO isn't resolving the cavity structure to the same level of detail as in the  C$^{18}$O PV diagram. \\

\noindent In Fig. \ref{PVD_13CO_cuts}, we display different cuts along the $^{13}$CO (3--2) PV diagram. We observe a velocity shift between the orange and purple velocity profiles of 0.45 km.s$^{-1}$ with the profile extracted via the orange profile (i.e. the profile extracted from the pseudo-slit that is further offset from the source position) at a velocity of V$_{lsr}$ = 7.41 km.s$^{-1}$ compared to V$_{lsr}$ = 6.96 km.s$^{-1}$ for the purple profile. This velocity shift is above the velocity resolution of the $^{13}$CO data (i.e. 0.221 km.s$^{-1}$; see Table \ref{beam_params_table}). Similarly, in Fig. \ref{PVD_C18O_cuts} (right panel) we also observe a velocity shift between the extracted velocity profiles from the C$^{18}$O PV diagram. A shift of 0.25 km.s$^{-1}$ was observed, with the velocity of the orange profile at 7.37 km.s$^{-1}$ compared to 7.12 km.s$^{-1}$ for the purple profile. The observed velocity shift larger than the velocity resolution of the C$^{18}$O data (i.e. 0.083 km.s$^{-1}$; see Table \ref{beam_params_table}), and aligns with the velocity shift observed in $^{13}$CO. An increase in the velocity of the material traced by both of these emission lines with increasing offset from the source position is indicative of slowly outflowing gas \citep[e.g. ][]{Ferreira1997, Pascucci2023} along a PA of 234$^{\circ}$. This is interesting for two reasons. Firstly, this emission is redshifted by > 2 km.s$^{-1}$ from the systemic velocity and secondly, is at a PA that has not been previously reported as the direction of a jet (e.g., a PA of 20$^{\circ}$ was reported by \cite{Podio2021} for the redshifted CO molecular jet emanating from IRAM04191) or outflow (PAs of 18$^{\circ}$ and 30${^\circ}$ have been proposed by \cite{CFLee2002,CFLee20052005} for the large scale CO outflow). The kinematic analysis presented here suggests that there is an outflow cavity to the south/south-west of IRAM04191. The detection of an outflow cavity at this PA then naturally leads to the idea that there must be an outflow present, separate to the previously detected molecular jets/outflows at other PAs to facilitate the generation of such a cavity. Given the above, this gives further credense to the work of \cite{Huelamo2026} that the IRAM04191 system could be a binary as there is a hint in our observations of a  outflow at small spatial scales.  
\begin{figure}[tbh!]
    \centering
    \includegraphics[width=0.75\linewidth]{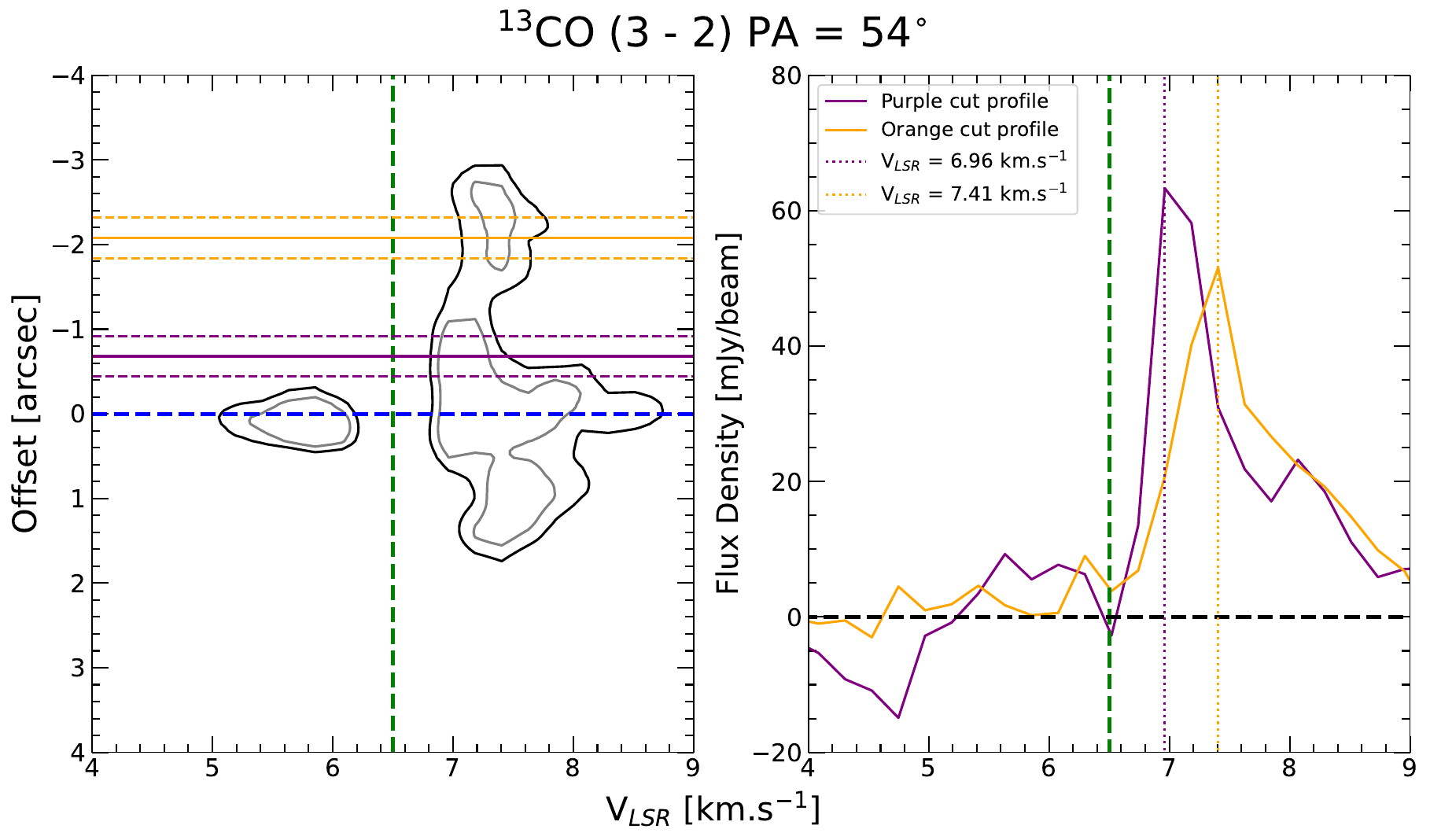}
    \caption{\textbf{Left:} The PV diagram for $^{13}$CO as shown in Fig. \ref{PVD_13co_vs_c18o_pa54}. Overlaid in orange and purple are the pseudo-slits (solid line = slit centre, dashed lines = slit edges) used to extract velocity profiles at these slit positions. The green and blue dashed lines correspond to the systemic velocity (Table \ref{properties table}) and zero offset from the source position respectively. \textbf{Right:} The velocity profiles extracted via the orange and purple pseudo-slits. These profiles are generated via median combination of the profiles along the width of each pseudo-slit. The orange and purple dotted lines indicate the velocity at which the peak flux density of each profile is observed. The green and black dashed lines correspond to the systemic velocity (Table \ref{properties table}) and zero flux density respectivily. }
    \label{PVD_13CO_cuts}
\end{figure}

\begin{figure}[tbh!]
    \centering
    \includegraphics[width=0.75\linewidth]{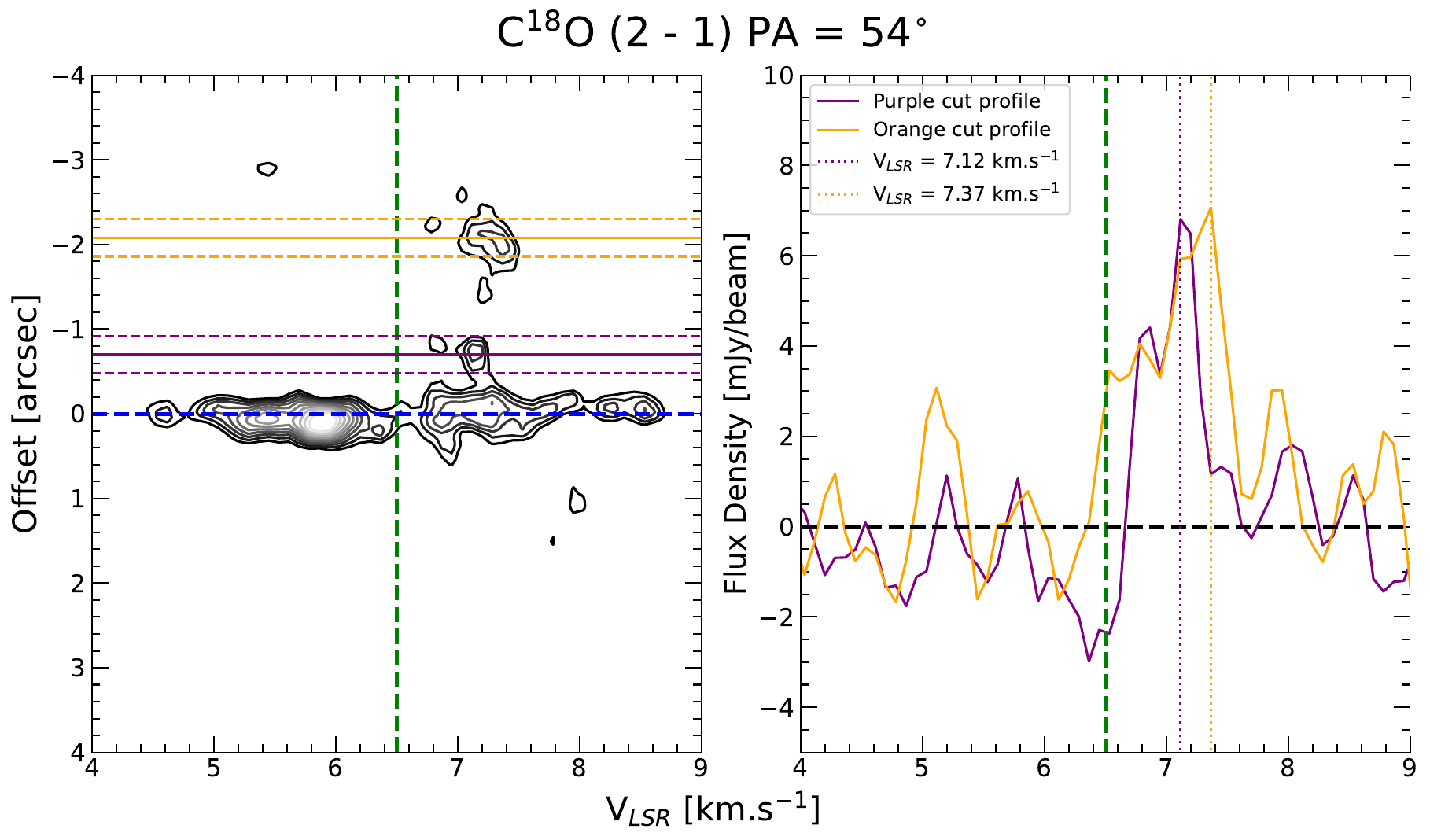}
    \caption{\textbf{Left:} The PV diagram for C$^{18}$O as shown in Fig. \ref{PVD_13co_vs_c18o_pa54}. Overlaid in orange and purple are the pseudo-slits (solid line = slit centre, dashed lines = slit edges) used to extract velocity profiles at these slit positions. The green and blue dashed lines correspond to the systemic velocity (Table \ref{properties table}) and zero offset from the source position respectively. \textbf{Right:} The velocity profiles extracted via the orange and purple pseudo-slits. These profiles are generated via median combination of the profiles along the width of each pseudo-slit. The orange and purple dotted lines indicate the velocity at which the peak flux density of each profile is observed. The green and black dashed lines correspond to the systemic velocity (Table \ref{properties table}) and zero flux density respectivily. }
    \label{PVD_C18O_cuts}
\end{figure}

\end{appendix}

\end{document}